\documentclass[a4paper,11pt]{article}

\usepackage{jheppub} 

\usepackage[T1]{fontenc} 

\usepackage{amsmath, amsfonts, amsthm, amssymb, graphicx, slashed}
\usepackage{subfig}
\usepackage{epstopdf}
\usepackage{mathrsfs}
\usepackage{enumerate}
\usepackage{hyperref}
\usepackage{color}

\title{\boldmath Disformally self-tuning gravity}

\author[]{William T. Emond}
\author[]{Paul M. Saffin}

\affiliation[]{School of Physics and Astronomy, 
University of Nottingham, University Park, Nottingham NG7 2RD, UK}

\emailAdd{william.emond@nottingham.ac.uk}
\emailAdd{paul.saffin@nottingham.ac.uk}

\abstract{We extend a previous self-tuning analysis of the most general scalar-tensor theory of gravity in four dimensions with second order field equations by considering a generalized coupling to the matter sector. Through allowing a disformal coupling to matter we are able to extend the {\it Fab Four} model and construct a new class of theories that are able to tune away the cosmological constant on Friedmann-Lema\^{i}tre-Robertson-Walker backgrounds.}


\begin{document} 
\maketitle
\flushbottom

\section{Introduction}
\label{sec:intro}

Over the past few years there has been a lot of interest in attempts to explain some of the current major issues of modern cosmology, such as \textit{dark energy} and \textit{dark matter}, 
through constructing modified theories of gravity. One particular obstacle that presents itself even before one considers such issues, producing severe complications for proposed explanations of dark 
energy, is the so-called ``\textit{cosmological constant problem}'' \cite{weinberg1989cosmological,Polchinski:2006gy,Burgess:2013ara,Martin:2012bt,Nobbenhuis:2004wn}.

The cosmological constant problem is one that arises through combining our knowledge from the two pillars of 20$^{th}$ century physics, quantum field theory (QFT) and general relativity (GR). Indeed, QFT predicts that the vacuum must have non-trivial structure (in particular, a non-zero energy), the source of which derives from the vacuum fluctuations of each quantum field. If we neglect gravitational effects (as in ``ordinary'' QFT) then this is not necessarily a problem \textit{per se}, since physically one cannot measure energy in an absolute sense, only 
relatively. However, taking gravity into account introduces a dilemma as gravity is sensitive to absolute energies, and as such, upon applying a regularisation procedure to calculate the energy contributions
from vacuum loop diagrams, it is found that this leads to a vacuum-energy contribution of order $M_{particle}^{4}$ for each particle species. This is problematic as it renders the vacuum energy divergent.
Naively, it is possible to fix this issue through introducing a \textit{bare} term $\Lambda_{0}$, which is itself divergent,  into Einstein's equation such that it counteracts the contribution from the particle content. In doing so one renormalises the cosmological constant such that what actually gravitates is the finite net cosmological constant,
\\
\begin{equation} \label{net cosmo const}
 \Lambda=\Lambda_{0} +\langle\rho_{m}\rangle_{vac}.
\end{equation}
\\
Current observational data requires that $\Lambda\sim (meV)^{4}$ \cite{Ade:2015xua} and so we see that a significant amount of fine-tuning is required in order for 
theory to match experiment.

This in itself is a problem, however, it is not catastrophic and it is not the real issue as to why the cosmological constant is so troubling. Indeed, this issue is far more acute than a fine-tuning problem, and arises from the fact that the renormalisation of the cosmological constant is not stable to radiative corrections. Higher order loop corrections, or changes in the matter sector, lead to large changes in $\Lambda$ at the scale of the QFT cut-off, meaning that $\Lambda_0$ is sensitive to high-energy physics, right up to the Planck scale. (For a detailed discussion of the cosmological constant problem we refer you to \cite{padilla2015lectures})
\\

\section{Self-tuning Horndeski theory \& a disformal coupling to matter}

Given that current experimental evidence is consistent with GR and Standard Model (SM) QFT, then any attempt to 
resolve this problem (without introducing any beyond the SM physics) must involve some form of modified theory of gravity. A particular approach to constructing a modified theory of gravity that has been employed often is one involving scalar-tensor combinations. 
Indeed this has proven to be a useful approach in a wide range of models, from Brans-Dicke gravity \cite{brans1961mach}, to more recent models 
\cite{deffayet2009covariant,deffayet2010imperfect,silva2009self,kobayashi2010inflation,de2011generalized,chiba2014cosmological,gubitosi2011purely,de2007kinetic,van2015disformally} inspired by Galileon theory \cite{nicolis2009galileon}. The starting point for us
is Horndeski's scalar-tensor theory of gravity \cite{horndeski1974second}, first discovered by \textit{G.W. Horndeski} in \textit{1974}, and independently re-discovered more recently by 
\textit{C. Deffayet et al.} \cite{deffayet2011k}. This is the most general scalar-tensor theory that produces second-order field equations, which is an essential requirement in order to avoid any 
Ostrogradski instability \cite{ostrogradskiǐ1850memoire} in the theory. Indeed, Horndeski theory has received a renewed interest in research into modified theories of gravity in recent years \cite{zumalacarregui2014transforming,anabalon2014asymptotically,kobayashi2011generalized,sotiriou2014black}.
We focus our attention on this particular approach, inspired by its effectiveness at providing a compelling proposal to rectify the cosmological constant problem in recent research. Indeed, such a solution was 
realised through the derivation of a class of self-tuning theories, the so-called \textit{Fab-Four} \cite{Charmousis:2011bf,charmousis2012self}, the argument here being that instead of concerning oneself over how to 
treat the vacuum energy contributions and radiative instability of the cosmological constant head-on, one can instead mitigate the effects of the cosmological constant on the geometry seen by matter, i.e. it simply does not gravitate. Preliminary analysis following the derivation of the Fab-Four also suggests that its radiative corrections can be kept under control  \cite{Charmousis:2011bf,charmousis2012self}, an essential requirement given the source of the cosmological constant problem in the first place. This ``\textit{screening}'' of the net cosmological constant from the gravitational sector (such that it is not a source of curvature) is achieved by constructing a theory of gravity, using the Horndeski action, that describes an interaction between gravity and some scalar field $\phi$. 

The basic idea of such a screening mechanism, as described by Weinberg \cite{weinberg1989cosmological}, is that the cosmological constant is absorbed by the scalar field's dynamics rather than the dynamics of gravity. Importantly, the Fab-Four avoids Weinberg's \textit{no-go theorem} \cite{weinberg1989cosmological} for self-tuning solutions by forgoing the requirement of Poincar\'{e} invariance of the self-tuning scalar field $\phi$. That is to say, $\phi$ is spatially homogeneous (satisfying the cosmological principle), but is allowed to evolve in time (i.e. $\phi =\phi (t)$) to enable it to \textit{``self-tune''} relative to the value of the vacuum energy at any given instant in time. In accordance with Einstein's equivalence principle, matter is then 
introduced into the theory by minimally coupling it to gravity via the metric $g_{\mu\nu}$, however, it is not coupled to the scalar field -  this interacts purely with gravity, acting as a screening mechanism to cloak the vacuum energy density $\langle\rho_{m}\rangle_{vac}$ (contributed from the matter sector) from the gravitational sector such that $\langle\rho_{m}\rangle_{vac}$ has no impact on the spacetime curvature. In doing so, the theory is described by the following action
\\
\begin{equation} \label{Fab-Four action}
S = S_{H}\left[g_{\mu\nu}, \phi\right] + S_{m}\left[g_{\mu\nu}, \psi_{i}\right] ,
\end{equation}
\\
where $S_{H}$ is the Horndeski action, $S_{m}$ is the effective action for matter, and $\psi_{i}$ are the matter fields, which are minimally coupled to $g_{\mu\nu}$. 
Using this approach, the Fab-Four was shown to be the most general self-tuning theory of gravity in which matter is minimally-coupled to gravity. 

Given that it has now been shown that a self-tuning theory as a solution to the cosmological constant problem is possible, it is natural to question whether it is feasible to construct a generalisation 
of this idea. Indeed, a minimal extension of the Fab-Four theory has recently been proposed by \emph{E. Babichev et al.} in which the starting point involves replacing the potentials appearing in the self-tuning Fab-Four Lagrangian with more general functions that depend upon both the scalar field $\phi$ and the corresponding canonical kinetic term $X=-\frac{1}{2}\partial_{\mu}\phi\partial^{\mu}\phi$ \cite{babichev2015beyond}. Nevertheless, in the quest to construct the most general self-tuning Horndeski theory, we further extend the Fab-Four model to encompass the case in which the scalar field  $\phi$ is allowed to enter the matter-sector, such that it interacts with matter directly. 
This can most readily be achieved through a \textit{disformal} coupling of matter to gravity \cite{bekenstein1993relation}. 

The aim, therefore, is to use the general idea of self-tuning, where a scalar field absorbs the effects of $\Lambda$, but in a more general context where the self-tuning scalar is allowed to couple directly to matter.
In such an approach, one considers two distinct (but related) geometries: one defining the geometry on which matter plays out its dynamics, and one describing 
gravitation. It was shown by Bekenstein \cite{bekenstein1993relation} that the most general relation between the physical and gravitational geometries, described by the two metrics 
$\bar{g}_{\mu\nu}$ and $g_{\mu\nu}$ respectively, involving a scalar field $\phi$ that adheres to the weak equivalence principle and causality, is given by the following \textit{disformal} transformation
\\
\begin{equation} \label{disformal transformation}
 \bar{g}_{\mu\nu}(x)= A^{2}(\phi ,X)\left[g_{\mu\nu}(x)+B^{2}(\phi ,X)\partial_{\mu}\phi\partial_{\nu}\phi\right],
\end{equation}
\\
where $X=-\frac{1}{2}g^{\mu\nu}\partial_{\mu}\phi\partial_{\nu}\phi$.
The inverse to \eqref{disformal transformation} is found to be \cite{zumalacarregui2013dbi}
\\
\begin{equation} \label{inverse disformal metric}
 \bar{g}^{\mu\nu}(x)=\frac{1}{A^{2}(\phi ,X)}\left[g^{\mu\nu}(x)-\frac{B^{2}(\phi ,X)}{1-2B^{2}(\phi ,X)X}g^{\mu\lambda}(x)g^{\nu\sigma}(x)\partial_{\lambda}\phi\partial_{\sigma}\phi\right],
\end{equation}
\\
which leads to an expression for $\bar{X}$ of the form,
\\
\begin{equation} \label{disformal X}
 \bar{X}(\phi,X)=-\frac{1}{2}\bar{g}^{\mu\nu}\partial_{\mu}\phi\partial_{\nu}\phi =\frac{X}{A^{2}(\phi ,X)\left(1-2B^{2}(\phi ,X)X\right)}.
\end{equation}
\\
In principle, this relation between $X$ and $\bar{X}$  can be inverted such that $X=X(\phi,\bar{X})$. With this in mind we can then rearrange 
\eqref{disformal transformation} to obtain an expression for the inverse transformation between the two metrics $g_{\mu\nu}$ and $\bar{g}_{\mu\nu}$, 
\\
\begin{align} \label{inverse disformal transformation}
 g_{\mu\nu}(x)&= \frac{1}{A^{2}(\phi ,X(\phi,\bar{X}))}\left[\bar{g}_{\mu\nu}(x)-A^{2}(\phi ,X(\phi,\bar{X}))B^{2}(\phi ,X(\phi,\bar{X}))\partial_{\mu}\phi\partial_{\nu}\phi\right]
\nonumber\\ \\ \nonumber 
&=\bar{A}^{2}(\phi ,\bar{X})\left[\bar{g}_{\mu\nu}(x)+\bar{B}^{2}(\phi ,\bar{X})\partial_{\mu}\phi\partial_{\nu}\phi\right],
\end{align}
\\
and from this we can imply that $\bar{A}$ and $\bar{B}$ are related to $A$ and $B$ in the following manner
\\
\begin{equation} \label{barA barB constraints}
 \bar{A}^{2}(\phi ,\bar{X})=\frac{1}{A^{2}(\phi ,X(\phi,\bar{X}))},\qquad \qquad \bar{B}^{2}(\phi ,\bar{X})=-A^{2}(\phi ,X(\phi,\bar{X}))B^{2}(\phi ,X(\phi,\bar{X})).
\end{equation}
\\

Given recent research into modified theories of gravity involving disformal couplings \cite{sakstein2014disformal,sakstein2015towards,koivisto2012screening,koivisto2008disformal,brax2014constraining}, 
it is natural to consider such an approach in an attempt to generalise the Fab-Four theory. Indeed, we aim to construct a self tuning theory of gravity described by the action 
\\
\begin{equation} \label{Einstein-frame action}
S = S_{H}\left[g_{\mu\nu}, \phi\right] + S_{m}\left[\bar{g}_{\mu\nu}, \psi_{i}\right].
\end{equation}
\\
In this representation of the theory both gravity \textit{and} matter are directly coupled to the scalar field $\phi (t)$ since there is a non-trivial dependence on $\phi$ contained in 
$\bar{g}_{\mu\nu}$. The reason being that we treat $g_{\mu\nu}$, $\phi$ and $\psi$ as the dynamical variables, with the physical geometry, $\bar{g}_{\mu\nu}$,  determined via \eqref{disformal transformation}, i.e. $\bar{g}_{\mu\nu}=\bar{g}_{\mu\nu}[g_{\mu\nu},\phi,\psi]$. We follow the convention of referring to these different representations of the theory as \textit{``frames''}, and shall refer to this particular representation as the 
\textit{Horndeski frame}, which is the analogue of the Einstein frame.

An alternative representation of the theory can be found by expressing the action in terms of $\bar{g}_{\mu\nu}$, $\phi$ and $\psi$, and using these as the dynamical variables. In doing so we remove any direct coupling 
of the scalar field $\phi$ to the matter-sector at the level of the action, but gravity remains directly coupled to $\phi$. We refer to this representation as the \textit{Jordan frame}, and in this frame the action can be expressed as 
\footnote{Note that $S_{J}$ will not in general be of Horndeski form, and is more likely to be of beyond-Horndeski form \cite{Zumalacarregui:2013pma,Gleyzes:2014dya}. This is because the Horndeski Lagrangian changes its form under general disformal transformations and so the gravitational action will no-longer resemble that of Horndeski theory in the Jordan frame.}
\begin{equation} \label{Jordan-frame action}
S = S_{J}\left[\bar{g}_{\mu\nu}, \phi\right] + S_{m}\left[\bar{g}_{\mu\nu}, \psi_{i}\right] .
\end{equation}

Writing the action in the Jordan frame is clearly advantageous as in this particular representation matter follows the geodesics defined by the physical metric $\bar{g}_{\mu\nu}$ 
(as opposed to in the Horndeski frame where the matter geodesics defined by the metric are also influenced by variations in the scalar field $\phi$), such that the associated energy-momentum tensor is covariantly 
conserved thus corresponding to the physical frame.

As the aim here is to construct a generalisation of the Fab-Four theory, in order to define the notion of \textit{self-tuning} we shall follow the structure laid out in 
the derivation of the Fab-Four theory \cite{Charmousis:2011bf,charmousis2012self}. Thus, by a self-tuning theory, we postulate that their exists some scalar field, $\phi$, that evolves dynamically such that it absorbs 
any energy density contributed by the net cosmological constant, and in doing so screens it from the gravitational sector such that it has no effect on the spacetime curvature that matter sees. In other words, the existence of this scalar field 
means that (effectively) the net cosmological constant (regardless of its value) does not gravitate. We do not require that such a scalar field is Poincar\'{e} invariant, and in doing so avoid 
Weinberg's \emph{no-go theorem} for such self-tuning modifications of gravity \cite{weinberg1989cosmological}.

In addition to this, the requirements that must be satisfied in order for the theory to be self-tuning, a so-called \textit{self-tuning filter} \cite{Charmousis:2011bf,charmousis2012self}, are as follows:
\\
\begin{enumerate}
 \item The vacuum solution, for the metric that matter sees ($\bar g_{\mu\nu}$), to such a theory should always be 
       Minkowski spacetime no matter the value of the net cosmological constant; \\

 \item This should remain so even after any phase transition in which the cosmological constant jumps instantaneously by a finite amount;\\

 \item The theory should permit a non-trivial cosmology (ensuring that Minkowski spacetime is not the only solution, a condition that is certainly required by
       observation). 
\\
\end{enumerate}
\label{self-tuning constraints}

\newpage

\section{Towards a self-tuning disformal theory of gravity}

In analogy to the derivation of the Fab-Four we first consider the Horndeski Lagrangian, given by 
\begin{equation} \label{Horndeski}
\mathscr{L}_{H}= \sum_{i=2}^{5}\mathscr{L}_{i}.
\end{equation}
Up to total derivative terms that do not contribute to the equations of motion, the different pieces can be written as \cite{zumalacarregui2013dbi}
\begin{align}\label{Lagrangian components}
\mathscr{L}_{2} &= K(\phi, X) ,
\\ \nonumber\\ 
\mathscr{L}_{3} &= -G_{3}(\phi, X)\Box\phi ,
\\ \nonumber\\ 
\mathscr{L}_{4} &= G_{4}(\phi, X)R+G_{4,X}\left[\left(\Box\phi\right)^{2}-\left(\nabla_{\mu}\nabla_{\nu}\phi\right)^{2}\right] ,
\\ \nonumber\\
\mathscr{L}_{5} &= G_{5}(\phi, X)G^{\mu\nu}\nabla_{\mu}\nabla_{\nu}\phi - \frac{1}{6} G_{5,X} \left[\left(\Box\phi\right)^{3} - 3 
\left(\Box\phi\right)\left(\nabla_{\mu}\nabla_{\nu}\phi\right)^{2} + 2 \left(\nabla_{\mu}\nabla_{\nu}\phi\right)^{3}\right] ,
\end{align}
where $G^{\mu\nu}$ is the Einstein tensor, $R$ the scalar curvature, $\Box\phi=\nabla^{\mu}\nabla_{\mu}\phi$, 
$\;\;\left(\nabla_{\mu}\nabla_{\nu}\phi\right)^{2}=\nabla^{\mu}\nabla^{\nu}\phi\nabla_{\mu}\nabla_{\nu}\phi$, 
$\left(\nabla_{\mu}\nabla_{\nu}\phi\right)^{3}=\nabla^{\mu}\nabla_{\nu}\phi\nabla^{\nu}\nabla_{\alpha}\phi\nabla^{\alpha}\nabla_{\mu}\phi\;\;$ and 
$\;\;X\equiv-\frac{1}{2}g^{\mu\nu}\partial_{\mu}\phi\partial_{\nu}\phi\;\;$ is the canonical kinetic term of the scalar field. Furthermore, note that $G_{i,X}\equiv\frac{\partial G_{i}}{\partial X}$. 

To enable its construction we first study the cosmological set-up of this theory. Indeed, we shall require that the geometry in both frames (Horndeski and Jordon, respectively) is Friedmann-Lema\^{i}tre-Robertson-Walker (FLRW). To this end, in the Horndeski frame, we treat $g_{\mu\nu}$, $\phi$ and $\psi_{i}$ as the dynamical variables, 
with the Jordan-frame metric $\bar{g}_{\mu\nu}$ being determined via \eqref{disformal transformation}. We then assume that the Horndeski-frame metric $g_{\mu\nu}$ abides by the cosmological principle, i.e. we require that at any given instant in time $t$ the geometry defined by $g_{\mu\nu}$ is spatially homogeneous and isotropic. 
This requirement is achieved by foliating spacetime into a set of spacelike hypersurfaces, $\Sigma_{t}$, such that the spatial ``slice'' at a given instant in time is homogeneous and isotropic. Accordingly the geometry defined by $g_{\mu\nu}$ is of the form, 
\\
\begin{equation} \label{gravitational line element}
ds^{2}=g_{\mu\nu}(x)dx^{\mu}dx^{\nu}=-N^{2}(t)dt^{2}+a^{2}(t)\gamma_{ij}(\mathbf{x})dx^{i}dx^{j},
\end{equation}
\\
where $N(t)$ is the \emph{lapse function}, $a(t)$ is the \emph{scale factor}, and $\gamma_{ij}(\mathbf{x})$ is the (maximally symmetric) metric on the plane ($k=0$), sphere ($k=1$), or hyperboloid 
($k=-1$).

In the Jordan frame, there is no direct interaction between the scalar field $\phi(t)$ and matter, instead $\bar{g}_{\mu\nu}$, $\phi$ and $\psi_{i}$ are treated as the  dynamical variables, with $g_{\mu\nu}$ determined via \eqref{inverse disformal transformation}. It seems reasonable to consider this as the physical frame since matter follows the geodesics defined purely by metric in this frame (as opposed to in the Horndeski frame where the matter geodesics defined by the metric are also influenced by variations in the scalar field $\phi$), and its 
corresponding energy-momentum tensor $\bar{T}^{\mu\nu}$ is locally conserved (i.e. $\bar{\nabla_{\mu}}\bar{T}^{\mu\nu}=0$, whereas in the Horndeski frame $\nabla_{\mu}T^{\mu\nu}\neq 0$ due to the direct coupling between $\phi$ and matter). Since we require that the matter sector is completely screened from any gravitational effects that would be introduced by the vacuum energy this amounts to stipulating that geometry defined by $\bar{g}_{\mu\nu}$ is asymptotically Minkowski in form,
\\
\begin{equation} \label{matter line element}
d\bar{s}^{2}=\bar{g}_{\mu\nu}(x)dx^{\mu}dx^{\nu}=-dt^{2}+\bar{a}^{2}(t)\gamma_{ij}(\mathbf{x})dx^{i}dx^{j}.
\end{equation}
\\
An immediate consequence of imposing the geometries \eqref{gravitational line element} and \eqref{matter line element}, along with the requirement that the scalar field is time-dependent 
(i.e. $\phi=\phi(t)$), is that the canonical kinetic term $X$ takes the form
\\
\begin{equation} \label{kinetic-term Einstein}
X=- \frac{1}{2}g^{00}\partial_{0}\phi\partial_{0}\phi = \frac{1}{2}\left(\frac{\dot{\phi}}{N}\right)^{2} ,
\end{equation} 
\\
when evaluated in the Horndeski frame, and 
\\
\begin{equation} \label{kinetic-term Jordan}
\bar{X}=- \frac{1}{2}\bar{g}^{00}\partial_{0}\phi\partial_{0}\phi = \frac{1}{2}\dot{\phi}^{2} ,
\end{equation} 
when evaluated in the Jordan frame.
We can also determine a mapping between the Horndeski and Jordan frames for the lapse function $N(t)$ and the scale factor $a(t)$ by noting that the geometry defined by $g_{\mu\nu}$ 
\eqref{gravitational line element} can be expressed in the Jordan frame via the inverse disformal transformation \eqref{inverse disformal transformation}
\begin{equation} \label{gravitational line element-Jordan frame}
ds^{2}= -\bar{A}^{2}(\phi ,\bar{X})\left[1-\bar{B}^{2}(\phi ,\bar{X})\dot{\phi}^{2}\right]dt^{2}
+\bar{a}^{2}(t)\bar{A}^{2}(\phi ,\bar{X})\left[\frac{dr^{2}}{1-kr^{2}}+r^{2}d\Omega^{2}\right].
\end{equation}
Now the two expressions \eqref{gravitational line element} and \eqref{gravitational line element-Jordan frame} must be equivalent, thus enabling us to infer the following relations
\begin{align} \label{transformed factors}
 N^{2}(t)&= \bar{A}^{2}(\phi ,\bar{X})\left[1-\bar{B}^{2}(\phi ,\bar{X})\dot{\phi}^{2}\right], \\ \nonumber\\
 a^{2}(t)&= \bar{a}^{2}(t)\bar{A}^{2}(\phi ,\bar{X}),
\end{align}
as well as confirming the earlier relation (\ref{disformal X}).

Finally, from the disformal relationship between the two metrics $g_{\mu\nu}$, $\bar{g}_{\mu\nu}$ \eqref{disformal transformation} and the form of $X$ \eqref{kinetic-term Einstein}, we can 
ascertain that the integration measures $\sqrt{-g}$ and $\sqrt{-\bar{g}}$ are related by \cite{zumalacarregui2013dbi,sakstein2015towards} 
\begin{equation} \label{integration measure}
\sqrt{-\bar{g}}=A^{4}\sqrt{1-2B^{2}X}\sqrt{-g}.
\end{equation}
Before continuing, a remark must be made about the cosmological set-up and its effect on the action for the theory. Since we are considering a homogeneous scalar field $\phi =\phi(t)$ it 
follows that the Lagrangian $\mathcal{L}_{H}(x)=\mathcal{L}_{H}(t)$ will also be homogeneous. As such we can effectively neglect spatial dependence in the theory in the following manner:
\begin{align}\label{redefined Horndeski action}
 S_{H}\left[g_{\mu\nu}, \phi\right]&=\int d^{4}x\;\sqrt{-g}\mathcal{L}_{H}(x)=\int dt\int d^{3}x\;N(t)a^{3}(t)\sqrt{\gamma}\mathcal{L}_{H}(t)
\nonumber\\ \nonumber \\&=\left(\int d^{3}x\;\sqrt{\gamma}\right)\int dt\;N(t)a^{3}(t)\mathcal{L}_{H}(t)
\nonumber\\ \nonumber\\ &\Rightarrow\qquad\tilde{S}_{H}\equiv\frac{S_{H}\left[g_{\mu\nu}, \phi\right]}{\int d^{3}x\;\sqrt{\gamma}}=\int dt\;N(\phi,X)a^{3}(t)\mathcal{L}_{H}(t)=\int dt\;\sqrt{-\tilde{g}}\mathcal{L}_{H}(t),
\end{align}
where we have used that $\sqrt{-g}=N(t)a^{3}(t)\sqrt{\gamma}$, and noted that $\gamma_{ij}(\mathbf{x})$ is a maximally symmetric spatial metric. Note that we have also factored out the spatial part of the determinant of the metric, as it plays no role. 
Accordingly, we define an \emph{effective} metric determinant $g\rightarrow\tilde{g}$ such that $\sqrt{-\tilde{g}}=N(t)a^{3}(t)=N(t)\bar{A}^{3}(\phi,\bar{X})\bar{a}^{3}(t)$. Having set-up the cosmological structure we are now in a position to evaluate the Horndeski Lagrangian \eqref{Horndeski} on a background FLRW cosmology. As the Horndeski 
Lagrangian provides a description of the dynamics in the gravitational sector, we evaluate the appropriate curvature terms on the metric defined in this sector. We then use the fact that the two 
metrics are disformally related to transform the pieces of (\ref{Horndeski}) to their expressions in terms of the physical metric via the inverse disformal transformation 
\eqref{inverse disformal transformation}.
After some work we obtain the following expression for the Horndeski Lagrangian \eqref{Horndeski} evaluated on a background FLRW cosmology
\newpage
\begin{equation}
\label{full l}
\begin{split}
\mathcal{L}_{_{FLRW}}&\equiv\sqrt{-\tilde{g}}\mathscr{L}_{H}\big\vert_{_{FLRW}}= \sqrt{-\tilde{g}}\sum_{i=2}^{5}\mathscr{L}_{i}\big\vert_{_{FLRW}}\\ \\
&=\bar{a}^{3}\Biggl[N\bar{A}^{3}K + \left(\bar{A}^{3}\right)^{\bullet}\frac{\dot{\phi}}{N}G_{3}-\dot{\phi}\tilde{G}_{3,\phi}+
6N\bar{A}G_{4}\frac{k}{\bar{a}^{2}}-6\frac{\bar{A}}{N}\dot{\bar{A}}^{2}G_{4}
-6\frac{\bar{A}^{2}}{N}\dot{\bar{A}}\dot{\phi}G_{4,\phi} \\ 
&\qquad\qquad+6\frac{\bar{A}}{N}\left(\frac{\dot{\phi}}{N}\right)^{2}\dot{\bar{A}}^{2}G_{4,X}+
\frac{3}{N}\dot{\bar{A}}\dot{\phi}G_{5}\frac{k}{\bar{a}^{2}}+\frac{\dot{\bar{A}}^{3}}{N^{2}}\left(\frac{\dot{\phi}}{N}\right)^{3}G_{5,X}
-3\dot{\phi}\tilde{G}_{5,\phi}\frac{k}{\bar{a}^{2}}-3\frac{\bar{A}}{N^{3}}\dot{\bar{A}}^{2}\dot{\phi}^{2}G_{5,\phi}\Biggr] \\ \\
&\qquad\quad+\bar{a}^{3}\Biggl[3\frac{\bar{A}^{3}}{N}\dot{\phi}G_{3}-3\tilde{G}_{3}+12\frac{\bar{A}^{2}}{N}\left(\frac{\dot{\phi}}{N}\right)^{2}\dot{\bar{A}}G_{4,X}
-6\frac{\bar{A}^{3}}{N}\dot{\phi}G_{4,\phi}-12\frac{\bar{A}^{2}\dot{\bar{A}}}{N}G_{4}\\
&\qquad\qquad\qquad+3\frac{\bar{A}}{N^{2}}\dot{\bar{A}}^{2}\left(\frac{\dot{\phi}}{N}\right)^{3}G_{5,X}
+3\frac{\bar{A}}{N}\dot{\phi}G_{5}\frac{k}{\bar{a}^{2}}-3\tilde{G}_{5}\frac{k}{\bar{a}^{2}}
-6\frac{\bar{A}^{2}}{N}\dot{\bar{A}}\left(\frac{\dot{\phi}}{N}\right)^{2}G_{5,\phi}\Biggr]\bar{H} \\ \\
&\qquad\quad+\bar{a}^{3}\Biggl[6\frac{\bar{A}^{3}}{N}\left(\frac{\dot{\phi}}{N}\right)^{2}G_{4,X}-6\frac{\bar{A}^{3}}{N}G_{4}
+3\frac{\bar{A}^{2}}{N^{2}}\dot{\bar{A}}\left(\frac{\dot{\phi}}{N}\right)^{3}G_{5,X}
-3\frac{\bar{A}^{3}}{N}\left(\frac{\dot{\phi}}{N}\right)^{2}G_{5,\phi}\Biggr]\bar{H}^{2} \\ \\
&\qquad\quad+\bar{a}^{3}\Biggl[\frac{\bar{A}^{3}}{N^{2}}\left(\frac{\dot{\phi}}{N}\right)^{3}G_{5,X}\Biggr]\bar{H}^{3} 
\end{split} 
\end{equation}
where $\;\bar{H}=\frac{\dot{\bar{a}}}{\bar{a}}$ is the Hubble parameter in the Jordan frame, with $\frac{d}{dt}(\;\;) \equiv \left(\;\;\right)^{\bullet}$, $\;(\;\;)_{,\phi}\equiv\frac{\partial}{\partial \phi}(\;\;)$ and $(\;\;)_{,X}\equiv\frac{\partial}{\partial X}(\;\;)$. Furthmore, we have defined to auxilliary functions $\tilde{G}_{3}$ and $\tilde{G}_{5}$, 
\\
\begin{align}\label{aux functions}
 \tilde{G}_{5,X}&\equiv\frac{\partial \tilde{G}_{5}}{\partial X}=\frac{N\bar{A}}{\dot{\phi}}G_{5}=\frac{\bar{A}}{\sqrt{2X}}G_5\nonumber\\ \\ 
 \tilde{G}_{3,X}&\equiv\frac{\partial \tilde{G}_{3}}{\partial X}=\frac{N\bar{A}^{3}}{\dot{\phi}}G_{3}=\frac{\bar{A}^3}{\sqrt{2X}}G_3 \nonumber
\end{align}
\\
It is then possible to express \eqref{full l} in the form 
\\
\begin{equation} \label{full l z form}
 \mathcal{L}_{_{FRW}}=\bar{a}^{3}\sum_{i=0}^{3}Z_{i}(\bar{a},\phi , \dot{\phi} , \ddot{\phi})\bar{H}^{i}.
\end{equation}
\\ 
As matter is decoupled from the scalar field $\phi$ (up to gravitational interactions) in the Jordan frame the on-shell equation of motion (EOM) for $\phi$ can be determined purely from \eqref{full l z form}. 
Indeed, from \eqref{full l z form} and using the generalised Euler-Lagrange equation it follows that 
\\
\begin{equation} \label{scalar eom}
\begin{split}
\varepsilon^{\phi}&= \frac{\partial\mathcal{L}}{\partial\phi} -\frac{d}{dt}\left[\frac{\partial \mathcal{L}}{\partial\dot{\phi}}\right]
+\frac{d^{2}}{dt^{2}}\left[\frac{\partial \mathcal{L}}{\partial\ddot{\phi}}\right], \\ \\
&= \bar{a}^{3}\sum_{i=0}^{3}\Biggl[\ddddot{\phi}Z_{i, \ddot{\phi} , \ddot{\phi}}\bar{H}^{i}
+\dddot{\phi}^{2}Z_{i, \ddot{\phi}, \ddot{\phi} , \ddot{\phi}}\bar{H}^{i}+\ddot{\phi}^{2}Z_{i, \dot{\phi}, \dot{\phi} , \ddot{\phi}}\bar{H}^{i}
+\dot{\phi}^{2}Z_{i, \phi, \phi , \ddot{\phi}}\bar{H}^{i}+\bar{a}^{2}Z_{i, \bar{a}, \bar{a} , \ddot{\phi}}\bar{H}^{i+2} \\ 
&\qquad\qquad\;\; +2\left(\dot{\phi}\dddot{\phi}Z_{i,\phi, \ddot{\phi} , \ddot{\phi}}+\ddot{\phi}\dddot{\phi}Z_{i,\dot{\phi}, \ddot{\phi} , \ddot{\phi}}
+\dot{\phi}\ddot{\phi}Z_{i,\phi, \dot{\phi} , \ddot{\phi}}\right)\bar{H}^{i}
+2\left(\dddot{\phi}Z_{i,\bar{a}, \ddot{\phi} , \ddot{\phi}}+\ddot{\phi}Z_{i,\bar{a}, \dot{\phi} , \ddot{\phi}}
+\dot{\phi}Z_{i,\bar{a} ,\phi, \ddot{\phi}}\right)\bar{H}^{i+1} \\ \\
&\qquad\qquad\;\; +2\left(3-i\right)\ddot{\phi}Z_{i, \dot{\phi} , \ddot{\phi}}\bar{H}^{i+1}
+2\left(3-i\right)\dddot{\phi}Z_{i, \ddot{\phi} , \ddot{\phi}}\bar{H}^{i+1}
+2\left(3-i\right)\dot{\phi}Z_{i, \phi , \ddot{\phi}}\bar{H}^{i+1} +\ddot{\phi}Z_{i,\bar{a} ,\phi, \ddot{\phi}}\bar{H}^{i}\\ \\
&\qquad\qquad\;\; +\dot{\phi}Z_{i,\phi, \dot{\phi}}\bar{H}^{i}-\ddot{\phi}Z_{i,\dot{\phi}, \dot{\phi}}\bar{H}^{i}
+2i\left(\dot{\phi}Z_{i,\phi, \ddot{\phi}}+\ddot{\phi}Z_{i,\dot{\phi}, \ddot{\phi}}
+\dddot{\phi}Z_{i, \ddot{\phi} , \ddot{\phi}}\right)\frac{\ddot{\bar{a}}}{\bar{a}}\bar{H}^{i-1}
+i\bar{a}Z_{i, \bar{a} ,\ddot{\phi}}\frac{\ddot{\bar{a}}}{\bar{a}}\bar{H}^{i} \\ \\
&\qquad\qquad\;\; +iZ_{i,\ddot{\phi}}\frac{\dddot{\bar{a}}}{\bar{a}}\bar{H}^{i-1}-iZ_{i,\ddot{\phi}}\ddot{\bar{a}}\bar{H}^{i}
+\left(i+1\right)\bar{a}Z_{i,\bar{a},\ddot{\phi}}\frac{\ddot{\bar{a}}}{\bar{a}}\bar{H}^{i}-\left(i+1\right)\bar{a}Z_{i,\bar{a},\ddot{\phi}}\bar{H}^{i+2}
+\bar{a}Z_{i,\bar{a},\ddot{\phi}}\bar{H}^{i+2} \\ \\
&\qquad\qquad\;\; +\left(3-i\right)\bar{a}Z_{i,\bar{a},\ddot{\phi}}\bar{H}^{i+2}+3iZ_{i,\ddot{\phi}}\frac{\ddot{\bar{a}}}{\bar{a}}\bar{H}^{i+1}
+\left(i+1\right)\left(3-i\right)Z_{i,\ddot{\phi}}\frac{\ddot{\bar{a}}}{\bar{a}}\bar{H}^{i}
-\left(i+1\right)\left(3-i\right)Z_{i,\ddot{\phi}}\frac{\ddot{\bar{a}}}{\bar{a}}\bar{H}^{i+2} \\ \\
&\qquad\qquad\;\; +\bar{a}Z_{i,\bar{a},\ddot{\phi}}\bar{H}^{i+2}-\left(3-i\right)Z_{i,\dot{\phi}}\bar{H}^{i+1}-\bar{a}Z_{i,\bar{a},\dot{\phi}}\bar{H}^{i+1}
-iZ_{i,\dot{\phi}}\frac{\ddot{\bar{a}}}{\bar{a}}\bar{H}^{i-1}+Z_{i,\phi}\bar{H}^{i}\Biggr].
\end{split}
\end{equation}
\\
And \textit{on-shell} this satisfies
\begin{equation} \label{on-shell scalar eom}
 \varepsilon^{\phi}=\varepsilon^{\phi}(\bar{a},\dot{\bar{a}},\ddot{\bar{a}},\phi,\dot{\phi},\ddot{\phi},\dddot{\phi},\ddddot{\phi})=0.
\end{equation}
\\
The Lagrangian \eqref{full l z form} can also be employed to calculate the gravitational Hamiltonian. Indeed, upon calculating the canonical kinetic momenta of each of the dynamical fields, $\bar{a}(t)$, $\phi(t)$ and $\dot{\phi}(t)$, 
(defined accordingly as $p_{_{\bar{a}}}\equiv\frac{\partial\mathcal{L}_{_{FRW}}}{\partial\bar{a}}$, $p_{_{\phi}}\equiv\frac{\partial\mathcal{L}_{_{FRW}}}{\partial\phi}$ and $p_{_{\dot{\phi}}}\equiv\frac{\partial\mathcal{L}_{_{FRW}}}{\partial\dot{\phi}}$ respectively), we find that  
\\
\begin{align}\label{Jordan frame Hamiltonian}
 \mathscr{H}&=\mathscr{H}(\bar{a},\dot{\bar{a}},\ddot{\bar{a}},\phi,\dot{\phi},\ddot{\phi},\dddot{\phi})
=\dot{\bar{a}}p_{_{\bar{a}}}+\dot{\phi}p_{_{\phi}}+\ddot{\phi}p_{_{\dot{\phi}}}-\mathcal{L}_{_{FRW}}\nonumber\\ \nonumber\\
&=\bar{a}^{3}\sum_{i=0}^{3}\biggl[\left((i-1)Z_{i}+\dot{\phi}Z_{i,\dot{\phi}}-\dot{\phi}^{2}Z_{i,\phi,\ddot{\phi}}-\dot{\phi}\ddot{\phi}Z_{i,\dot{\phi},\ddot{\phi}}+\ddot{\phi}Z_{i,\ddot{\phi}}
-\dot{\phi}\dddot{\phi}Z_{i,\ddot{\phi},\ddot{\phi}}\right)\bar{H}^{i}\nonumber\\ &\qquad\qquad\qquad +\left((i-3)\dot{\phi}Z_{i,\ddot{\phi}}
-\bar{a}\dot{\phi}Z_{i,\bar{a},\ddot{\phi}}\right)\bar{H}^{i+1}-i\dot{\phi}\frac{\ddot{\bar{a}}}{\bar{a}}Z_{i,\ddot{\phi}}\bar{H}^{i-1}\biggr],
\end{align}
\\
where, for example, $Z_{i,\dot{\phi}}\equiv\frac{\partial Z_{i}}{\partial\dot{\phi}}$ and $Z_{i,\phi,\ddot{\phi}}\equiv\frac{\partial^{2} Z_{i}}{\partial\ddot{\phi}\partial\phi}$ 
(the other derivative terms are defined in the same manner).

Accordingly, the full Hamiltonian $\mathscr{H}_{total}$ of the theory can be constructed from the contribution from the gravitational-sector $\mathscr{H}$ and a source from the matter-
sector in the form of a cosmological fluid of energy density $\rho_{m}$. Consequently, the full Hamiltonian is given by $\mathscr{H}_{total}=\mathscr{H}+\rho_{m}$. Importantly, as a result of the 
diffeomorphism invariance of the theory, the full Hamiltionian satisfies the constraint $\mathscr{H}_{total}=0$, such that 
\\
\begin{equation} \label{Hamiltonian constraint}
\mathscr{H} + \rho_{m} = 0 \quad\Rightarrow\quad \mathscr{H} = -\rho_{m} .
\end{equation}
\\

\section{Applying the self-tuning filter}

Up to this point we have kept things fairly general, only applying minimal constraints to our theory, however we now wish to pass it through our self-tuning filter (laid out at the end of section \ref{self-tuning constraints}) and determine the 
contraints that we must apply in order for the theory to be self-tuning. We shall apply the filter in the Jordan frame in which the spacetime geometry is described by $\bar{g}_{\mu\nu}$. The reasoning being that this is the metric to which matter couples, and as such is required to be screened from the effects of the cosmological constant, i.e. we want the theory to self-tune with respect to this metric.

We first consider the implications of applying the self-tuning filter in the situation where our cosmological background is in vacuo. Now, the matter-sector is expected to contribute a constant vacuum energy density that (as in the case of the Fab-Four derivation \cite{Charmousis:2011bf,charmousis2012self}) we identify with the cosmological constant, 
$\langle\rho_{m}\rangle=\Lambda$. According to the first filter the vacuum energy should have no effect on the spacetime curvature seen by matter, thus we require a flat spacetime regardless the 
value of $\Lambda$. It also follows from the second filter that this should remain true even in the case where the matter-sector undergoes a phase transition and in doing so alters the the net value 
of $\Lambda$ by a constant amount (over an effectively infinitesimal time interval). This translates to the requirement that any abrupt change in the matter-sector is completely absorbed by the scalar 
field $\phi$, leaving the geometry unaffected. Consequently the scalar field tunes itself to each change in the vacuum energy $\Lambda$ and this must be permitted independently of the time of 
transition.

Given these considerations, our initial observation is that in order to be consistent with the first filter we seek cosmological vacuum solutions that are Ricci flat i.e. $R=0$. Insisting on this provides us with 
so-called ``\emph{on-shell-in-$\bar{a}$}'' conditions
\\
\begin{align}\label{on-shell-in-a}
\bar{H}^{2}&=-\frac{k}{\bar{a}^{2}}\;\;\Rightarrow\;\;\dot{\bar{a}}=\sqrt{-k}, \\ \nonumber\\
\ddot{\bar{a}}&=0,
\end{align}
\\
where we have defined $s\equiv\sqrt{-k}$ such that $\bar{H}=\frac{s^2}{\bar{a}^{2}}$ when \emph{on-shell-in-$\bar{a}$}.

To proceed, we shall assume that the scalar $\phi$ is a continuous function, but that $\dot{\phi}$, $\ddot{\phi}$ and $\dddot{\phi}$ may be discontinuous. With this in mind, we then go 
\textit{on-shell-in-$\bar{a}$} at the level of the field equations. This means that we impose the conditions \eqref{on-shell-in-a}, but leave $\phi$ to be determined dynamically. 

In doing so we find that 
\\
\begin{align}
\mathscr{H}(\bar{a},\dot{\bar{a}},\ddot{\bar{a}},\phi,\dot{\phi},\ddot{\phi},\dddot{\phi})&\quad\rightarrow\quad\mathscr{H}_{k}(\bar{a}_{k},\phi,\dot{\phi},\ddot{\phi},\dddot{\phi}),\nonumber 
\\ \\
\varepsilon^{\phi}=\varepsilon^{\phi}(\bar{a},\dot{\bar{a}},\ddot{\bar{a}},\phi,\dot{\phi},\ddot{\phi},\dddot{\phi},\ddddot{\phi})&\quad\rightarrow\quad
\varepsilon^{\phi}_{k}(\bar{a}_{k},\phi,\dot{\phi},\ddot{\phi},\dddot{\phi},\ddddot{\phi}),\nonumber
\end{align}
\\
such that the \textit{on-shell-in-$\bar{a}$} field equations are
\\
\begin{equation}
\label{eq:hamiltonianLambda}
\mathscr{H}_{k}=-\Lambda,\qquad\varepsilon_{k}^{\phi}=0,
\end{equation}
\\
where, to adhere to the second filter, the matter sector contributes $\Lambda$ to the vacuum energy, where $\Lambda$ is a piece-wise constant function of time. A subscript/superscript $k$ on a variable will denote that it is \emph{on-shell-in-$\bar{a}$}.

From \eqref{Jordan frame Hamiltonian} it can be seen that the gravitational Hamiltonian $\mathscr{H}$ is constructed from a set of functions $Z_{i}=Z_{i}(\bar{a},\phi,\dot{\phi},\ddot{\phi})$ (and 
their derivatives), with terms  depending on $\dot{\phi}$, $\ddot{\phi}$ and $\dddot{\phi}$. As such, requiring that it satisfy the condition  \eqref{eq:hamiltonianLambda}  imposes retrictions on how $\ddot{\phi}$, $\dddot{\phi}$ appear in $\mathscr{H}$ (as $\mathscr{H}_{k}=-\Lambda$ there must be some discontinuity in $\mathscr{H}_{k}$ to account for the discontinuous nature of $\Lambda$). Noting also 
from \eqref{scalar eom} that the scalar EOM is similarly constructed from the set of functions $Z_{i}=Z_{i}(\bar{a},\phi,\dot{\phi},\ddot{\phi})$ (and their derivatives), we can use these restrictions 
to impose constraints on the functional form of $\varepsilon^{\phi}_{k}$. This leaves us with three possible cases to consider:
\\
\begin{enumerate}[i)] \label{Zi cases 1}
\item $Z^{k}_{i}(\bar{a},\phi,\dot{\phi},\ddot{\phi})$ is \emph{non-linear} in $\ddot{\phi}$;\\
\item $Z^{k}_{i}(\bar{a},\phi,\dot{\phi},\ddot{\phi})$ is \emph{linear} in $\ddot{\phi}$;\\
\item $Z^{k}_{i}(\bar{a},\phi,\dot{\phi},\ddot{\phi})$ is \emph{independent} of $\ddot{\phi}$.\\
\end{enumerate}
Indeed, requiring that $\mathscr{H}_{k}$ contains a discontinuity, accounting for the discontinuous nature of $\Lambda$, imposes the following contraints, which may be seen from  \eqref{Jordan frame Hamiltonian}:
\\
\begin{enumerate}[i)] \label{Zi cases 2}
\item If $Z^{k}_{i}(\bar{a},\phi,\dot{\phi},\ddot{\phi})$ is \emph{non-linear} in $\ddot{\phi}$ we require that $\dddot{\phi}\sim$ step-function which further implies that $\ddddot{\phi}\sim$ delta-
function;\\
\item If $Z^{k}_{i}(\bar{a},\phi,\dot{\phi},\ddot{\phi})$ is \emph{linear} in $\ddot{\phi}$ then we require that $\ddot{\phi}\sim$ step-function which then implies that $\dddot{\phi}\sim$ delta-function. 
Note also that, from \eqref{Jordan frame Hamiltonian}, this automatically implies that $\mathscr{H}_{k}$ is \emph{linear} in $\ddot{\phi}$;\\
\item Finally, if $Z^{k}_{i}(\bar{a},\phi,\dot{\phi},\ddot{\phi})$ is \emph{independent} of $\ddot{\phi}$, then $\dot{\phi}\sim$ step-function implying that $\ddot{\phi}\sim$ delta-function and $\mathscr{H}_{k}$ must be independent of $\ddot{\phi}$ and $\dddot{\phi}$. \\
\end{enumerate}
Following on from this analysis, we can then study the implications of these results on the equations of motion for the scalar field $\varepsilon^{\phi}_{k}$. Again, working on a case-by-case 
basis, and using the expression $\varepsilon^{\phi}_{k}$ \eqref{scalar eom} we find the following:
\\
\begin{enumerate}[i)] \label{Zi cases 3}
\item If $Z^{k}_{i}(\bar{a},\phi,\dot{\phi},\ddot{\phi})$ is \emph{non-linear} in $\ddot{\phi}$ this implies that, in general, $Z^{k}_{i,\ddot{\phi},\ddot{\phi}}\neq 0$ and accordingly $\varepsilon^{\phi}_{k}$ is, at 
most, \emph{linear} in $\ddddot{\phi}$. However, we require that $\varepsilon^{\phi}_{k}=0$ and noting that in this case $\ddddot{\phi}\sim$ delta-function, we must therefore conclude that 
$Z^{k}_{i,\ddot{\phi},\ddot{\phi}}=0$ (since there is no support for a delta-function on the left-hand side of the equation);\\
\item If $Z^{k}_{i}(\bar{a},\phi,\dot{\phi},\ddot{\phi})$ is \emph{linear} in $\ddot{\phi}$ then clearly $Z^{k}_{i,\ddot{\phi},\ddot{\phi}}= 0$ and it follows that $\varepsilon^{\phi}_{k}$ will be, at 
most, \emph{linear} in $\ddot{\phi}$ (note that $Z^{k}_{i,\dot{\phi},\dot{\phi}}\sim\alpha(\bar{a},\phi,\dot{\phi})\ddot{\phi}$, since $Z^{k}_{i}$ is \emph{linear} in $\ddot{\phi}$, and hence 
$Z^{k}_{i,\dot{\phi},\dot{\phi},\ddot{\phi}}\sim\alpha(\bar{a},\phi,\dot{\phi})\Rightarrow\ddot{\phi}^{2}Z^{k}_{i,\dot{\phi},\dot{\phi},\ddot{\phi}}\sim\ddot{\phi}^{2}\alpha(\bar{a},\phi,\dot{\phi})$. 
Accordingly, we see that this term will cancel with the term $-\ddot{\phi}Z^{k}_{i,\dot{\phi},\dot{\phi}}\sim\ddot{\phi}^{2}\alpha(\bar{a},\phi,\dot{\phi})$ in \eqref{Jordan frame Hamiltonian}. As such, any non-linear 
terms in $\ddot{\phi}$  in the Hamiltonian will cancel out);\\
\item If $Z^{k}_{i}(\bar{a},\phi,\dot{\phi},\ddot{\phi})$ is \emph{independent} of $\ddot{\phi}$ it is trivially found that $\varepsilon^{\phi}_{k}$ will be, at most, \emph{linear} in $\ddot{\phi}$.\\
\end{enumerate}
Hence, it is seen that in all three cases $Z^{k}_{i}$ can be, at most, \textit{linear} in $\ddot{\phi}$ and consequently the \textit{on-shell-in-$\bar{a}$} Lagrangian $\mathcal{L}_{k}$ will be also.

Following this analysis to its logical conclusion it is found that, in actual fact, in order to satisfy \eqref{on-shell-in-a} the \emph{on-shell-in-$\bar{a}$} Lagrangian must be 
equivalent to a total derivative (a detailed discussion on this analysis is provided in appendix A).

We are now in the position to construct a preliminary definition for a self-tuning Lagrangian that satisfies the \emph{on-shell-in-$\bar{a}$} condition \eqref{on-shell-in-a}. To this end, we take into 
account that two Lagrangians that differ by a total derivative describe the same dynamical theory (i.e. they lead to the same equations of motion). Thus, we are working within an equivalence class of 
Lagrangians, $\left[\mathcal{L},\equiv\right]$, where two Lagrangians are considered equivalent if (and only if) they differ by a total derivative
\\
\begin{equation}
\tilde{\mathcal{L}}\equiv\mathcal{L}\quad\iff\quad \tilde{\mathcal{L}}= \mathcal{L} + \frac{df}{dt}.
\end{equation}
\\

We further note that when \emph{on-shell-in-$\bar{a}$}, $\bar{H}=\frac{s}{\bar{a}}$, hence
\\
\begin{equation}
\mathcal{L}\quad\rightarrow\quad\mathcal{L}_{k}=\bar{a}^{3}\sum_{i=0}^{3}Z_{i}\left(\frac{s}{\bar{a}}\right)^{i}.
\end{equation}
\\
From our earlier analysis we also know that the \emph{on-shell-in-$\bar{a}$} Lagrangian must be equal to a total derivative
\\
\begin{equation}
\mathcal{L}_{k}=\bar{a}^{3}\sum_{i=0}^{3}Z_{i}\left(\frac{s}{\bar{a}}\right)^{i}= \small{total\;derivative}.
\end{equation}
\\
As such, we can construct the following \textit{``Horndeski-like''} Lagrangian
\\
\begin{align} 
\label{self-tuning Lagrangian}
\tilde{\mathcal{L}}&= \bar{a}^{3}\sum_{i=0}^{3}\tilde{Z}_{i}\bar{H}^{i} \equiv 
-\bar{a}^{3}\sum_{i=0}^{3}\tilde{Z}_{i}\left(\frac{s}{\bar{a}}\right)^{i}+\bar{a}^{3}\sum_{i=0}^{3}\tilde{Z}_{i}\bar{H}^{i} \nonumber \\ \nonumber\\
&= \bar{a}^{3}\sum_{i=1}^{3}\tilde{Z}_{i}\biggl[\bar{H}^{i}-\left(\frac{s}{\bar{a}}\right)^{i}\biggr].
\end{align}
\\
where $\tilde{Z}_{i}=\tilde{Z}_{i}(\bar{a},\phi,\dot{\phi},\ddot{\phi})$.

Such a Lagrangian certainly adheres to the self-tuning criteria (\emph{c.f.} end of section \ref{self-tuning constraints}); it is, in a sense, \emph{sufficient} for self-tuning, but to what extent is it \emph{necessary}? Indeed, a priori, it cannot be taken to 
be necessary as there could possibly be other equivalent Lagrangians, with $Z_{i}=\tilde{Z}_{i}+\Delta Z_{i}$, that admit the same set of self-tuning solutions. To establish whether this is the case we 
need to demand that the ``tilded'' and ``untilded'' systems each have equations of motion that give the same dynamics when on-shell (generically, i.e. not just when \emph{on-shell-in-$\bar{a}$}). That 
is, we require that when on-shell
\\
\begin{equation}
\label{generic-on-shell}
\mathscr{H}=-\rho_{m}\;,\quad\varepsilon^{\phi}=0\qquad\iff\qquad\tilde{\mathscr{H}}=-\rho_{m}\;,\quad\tilde{\varepsilon}^{\phi}=0 .
\end{equation}
\\
In general we cannot imply from this statement that $\varepsilon^{\phi}\equiv\tilde{\varepsilon}^{\phi}$, nor even that $\varepsilon^{\phi}\propto\tilde{\varepsilon}^{\phi}$, as there could 
well be a non-linear relation between all the relevant equations. Despite this it turns out that, in order for a general \textit{``Horndeski-like''} self-tuning theory to be viable, we are forced to 
have
\\
\begin{equation} \label{general self-tuning}
\mathscr{H}=\tilde{\mathscr{H}}\;,\quad \varepsilon^{\phi}=\tilde{\varepsilon}^{\phi}.
\end{equation}
\\
In other words our putative self-tuning Lagrangian $\tilde{\mathcal{L}}$ describes the general case of a self-tuning theory, satisfying the self-tuning constraints of section
\ref{self-tuning constraints}. Furthermore, as this result implies that $\Delta Z_{i}=0$ we find that $Z_{i}=\tilde{Z}_{i}$ and that these functions can at most be dependent on $\bar{a}$, $\phi$ and 
$\dot{\phi}$, i.e. $Z_{i}=Z_{i}(\bar{a},\phi,\dot{\phi})$. (Refer to appendix A for a detailed discussion on this analysis).

Given the functional expression for the Lagrangian \eqref{full l} (when evaluated on an FRW background) we observe that the functions $Z_{i}$ may be expressed in the following form
\\
\begin{equation}
\label{X and Y}
Z_{i}(\bar{a},\phi,\dot{\phi})=X_{i}(\phi,\dot{\phi})-\frac{k}{\bar{a}^{2}}Y_{i}(\phi,\dot{\phi})=X_{i}(\phi,\dot{\phi})+\left(\frac{s}{\bar{a}}\right)^{2}Y_{i} (\phi,\dot{\phi}).
\end{equation}
\\
and importantly, by comparison with \eqref{full l}, it is clear that $Y_{2}=Y_{3}=0$.

It is then possible to derive a set of equations relating the remaining non-trivial functions $X_{i}$ ($i=0,\ldots ,3$) and $Y_{i}$ ($i=0,1$). Indeed, from the analysis above we know that 
$\mathcal{L}\equiv\tilde{\mathcal{L}}$ and as such they differ by a total derivative at most. Taking this into account we find the following set of equations for $X_{i}$ and 
$Y_{i}$
\\
\begin{align}
\label{X,Y relations 1}
X_{0}&= -\Lambda + \dot{\phi}V'_{0}\left(\phi\right),\nonumber\\ \nonumber\\
X_{1}&=\dot{\phi}V'_{1}\left(\phi\right)+3V_{0}\left(\phi\right),\nonumber\\ \\
X_{2}+Y_{0}&=\dot{\phi}V'_{2}\left(\phi\right)+2V_{1}\left(\phi\right),\nonumber\\ \nonumber\\
X_{3}+Y_{1}&=\dot{\phi}V'_{3}\left(\phi\right)+V_{2}\left(\phi\right),\nonumber
\end{align}
\\
where $V_{i}=V_{i}(\phi)$ are arbitrary potential terms. (Refer to appendix B for a detailed discussion).

We can also derive a set of equations that relate $X_{i}$ and $Y_{i}$ to the component functions $K=K(\phi,X)$, $G_{i}=G_{i}(\phi,X)$ ($i=3,4,5$) of the Horndeski Lagrangian when evaluated on an FLRW 
background \eqref{full l} by direct comparison of \eqref{X and Y} with \eqref{full l} 
\\
\begin{align}
\label{X,Y relations 2}
X_{0}&= N\bar{A}^{3}K + \left(\bar{A}^{3}\right)^{\bullet}\dot{\phi}\frac{G_{3}}{N}-\dot{\phi}\tilde{G}_{3,\phi}-6\frac{\bar{A}}{N}\dot{\bar{A}}^{2}G_{4}
-6\frac{\bar{A}^{2}}{N}\dot{\bar{A}}\dot{\phi}G_{4,\phi}\nonumber\\
&\qquad+6\frac{\bar{A}}{N}\dot{\bar{A}}^{2}\left(\frac{\dot{\phi}}{N}\right)^{2}G_{4,X}
+\frac{\dot{\bar{A}}^{3}}{N^{2}}\left(\frac{\dot{\phi}}{N}\right)^{3}G_{5,X}
-3\frac{\bar{A}}{N^{3}}\dot{\bar{A}}^{2}\dot{\phi}^{2}G_{5,\phi},\nonumber\\ \nonumber\\
X_{1}&=3\frac{\bar{A}^{3}}{N}\dot{\phi}G_{3}-3\tilde{G}_{3}+12\frac{\bar{A}^{2}}{N}\left(\frac{\dot{\phi}}{N}\right)^{2}\dot{\bar{A}}G_{4,X}
-6\frac{\bar{A}^{3}}{N}\dot{\phi}G_{4,\phi}-12\frac{\bar{A}^{2}\dot{\bar{A}}}{N}G_{4}\nonumber\\
&\qquad+3\frac{\bar{A}}{N^{2}}\dot{\bar{A}}^{2}\left(\frac{\dot{\phi}}{N}\right)^{3}G_{5,X}
-6\frac{\bar{A}^{2}}{N}\dot{\bar{A}}\left(\frac{\dot{\phi}}{N}\right)^{2}G_{5,\phi},\nonumber\\ \\
X_{2}+Y_{0}&= 6\frac{\bar{A}^{3}}{N}\left(\frac{\dot{\phi}}{N}\right)^{2}G_{4,X}-6\frac{\bar{A}^{3}G_{4}}{N}
+3\frac{\bar{A}^{2}}{N^{2}}\dot{\bar{A}}\left(\frac{\dot{\phi}}{N}\right)^{3}G_{5,X}
-3\frac{\bar{A}^{3}}{N}\left(\frac{\dot{\phi}}{N}\right)^{2}G_{5,\phi}\nonumber\\&\qquad-6N\bar{A}G_{4}-
\frac{3}{N}\dot{\bar{A}}\dot{\phi}G_{5}+3\dot{\phi}\tilde{G}_{5,\phi},\nonumber\\ \nonumber\\
X_{3}+Y_{1}&=\frac{\bar{A}^{3}}{N^{2}}\left(\frac{\dot{\phi}}{N}\right)^{3}G_{5,X}-3\frac{\bar{A}}{N}\dot{\phi}G_{5}+3\tilde{G}_{5}.\nonumber
\end{align}
\\
Thus through equating the corresponding equations in \eqref{X,Y relations 1} and \eqref{X,Y relations 2} we obtain a set of partial differential equations in $X$ that the component functions $K(\phi,X)$ and $G_{i}(\phi,X)$ 
($i=3,4,5$) must satisfy in order for the theory to be self-tuning.
\\

\subsection{Recovering the Fab-Four}
Having determined the self-tuning contraints we now wish to begin analysing particular cases. An important first step is checking the consistency of 
the theory, i.e. that in the special case where $\bar{A}=1$, $N=1$ and $B=0$ it reduces to the Fab-Four. In this case the set of differential equations given in \eqref{X,Y relations 2} take the form
\\
\begin{align}
\label{Fab-Four X,Y relations}
X_{0}&= K-\dot{\phi}\tilde{G}_{3,\phi},\nonumber\\ \nonumber\\
X_{1}&=3\dot{\phi}G_{3}-3\tilde{G}_{3}-6\dot{\phi}G_{4,\phi},\nonumber\\ \\
X_{2}+Y_{0}&= 6\dot{\phi}^{2}G_{4,X}-12G_{4}-3\dot{\phi}^{2}G_{5,\phi}+3\dot{\phi}\tilde{G}_{5,\phi},\nonumber\\ \nonumber\\
X_{3}+Y_{1}&=\dot{\phi}^{3}G_{5,X}-3\dot{\phi}G_{5}+3\tilde{G}_{5}.\nonumber
\end{align} 
\\
Utilising the equations given in \eqref{X,Y relations 1} we can solve these iteratively, starting with $X_{3}+Y_{1}$ and rewriting it purely in terms of $\tilde{G}_5$. This then gives us a differential 
equation for $\tilde{G}_5$ which we can solve and subsequently deduce $G_5$ (using that $\tilde{G}_{5,X}=\frac{N\bar{A}}{\dot{\phi}}G_{5}$). Then, by inserting our solutions for $\tilde{G}_5$ and $G_5$ 
into the next equation in \eqref{Fab-Four X,Y relations}, $X_{2}+Y_{0}$, we can solve for $G_4$; continuing in this fashion we can also determine $\tilde{G}_3$, $G_{3}$ and finally $K$.

Accordingly, the following set of solutions are obtained:
\\
\begin{align} \label{Fab-Four solutions}
K(\phi, X)&= \text{const.}+\dot{\phi}^{2}f'_{3}(\phi)+\dot{\phi}^{4}f''_{4}(\phi)-\frac{1}{4}\dot{\phi}^{6}g'''_{5}(\phi)
-\frac{1}{8}\dot{\phi}^{4}\left[2\ln (\dot{\phi})-1\right]V''''_{3}(\phi), \nonumber\\ \nonumber\\
G_{3}(\phi, X)&=f_{3}(\phi)-\frac{1}{8}\dot{\phi}^{2}\left[6\ln (\dot{\phi})-\right]V'''_{3}(\phi)
+ 3\dot{\phi}^{2}f'_{4}(\phi)+\frac{5}{4}\dot{\phi}^{4}g''_{5}(\phi),\nonumber\\ \nonumber\\
\tilde{G}_{3}(\phi, X)&=-V_{0}+\dot{\phi}f_{3}(\phi)-\frac{1}{8}\dot{\phi}^{3}\left[2\ln (\dot{\phi})-1\right]V'''_{3}(\phi)+ \dot{\phi}^{3}f'_{4}(\phi)
+\frac{1}{4}\dot{\phi}^{5}g''_{5}(\phi),\nonumber\\ \\
G_{4}(\phi, X)&=\dot{\phi}^{2}f_{4}(\phi)-\frac{1}{4}\dot{\phi}^{2}\ln (\dot{\phi})V''_{3}(\phi)+\frac{1}{2}\dot{\phi}^{4}g'_{5}(\phi)-\frac{1}{6}V_{1}(\phi), \nonumber\\ \nonumber\\
G_{5}(\phi, X)&=f_{5}(\phi) +3\dot{\phi}^{2}g_{5}(\phi)-\frac{1}{2}\left[\ln (\dot{\phi})+1\right]V'_{3}(\phi),\nonumber\\ \nonumber\\
\tilde{G}_{5}(\phi, X)&=\dot{\phi}f_{5}(\phi) +\dot{\phi}^{3}g_{5}(\phi)-\frac{1}{2}\dot{\phi}\ln (\dot{\phi})V'_{3}(\phi)+\frac{1}{3}V_{2}(\phi), \nonumber
\end{align}
\\
where $X=\frac{1}{2}\dot{\phi}^{2}$. (Note that we can identify the constant in the expression for $K$ with the vacuum energy, such that $\text{const.}=-\Lambda$).

In this case we can follow a similar analysis as in the Fab-Four analysis \cite{Charmousis:2011bf,charmousis2012self} to deduce the covariant form of the self-tuning Lagrangian. Indeed, starting from the Horndeski 
Lagrangian evaluated on an FLRW background \eqref{full l} we note that each of the arbitrary potential terms $V_{i}$ ($i=0,\ldots, 3$) and integration functions $f_{3}$, $f_{4}$, $f_{5}$ and $g_{5}$ are 
completely de-coupled from one another. As such we can analyse the form of \eqref{full l} on a case-by-case basis and in doing so we find that the functions $V_0$, $V_2$, $f_3$  lead to vanishing contributions in \eqref{full l}; $V_1$, $V_3$, $f_4$, $f_5$ and $g_5$ give non trivial terms, but $f_4$ and $f_5$ lead to the same type of expression, meaning that only four of the eight functions yield independent terms in the self-tuning FLRW Lagrangian. Upon lifting this from the FLRW Lagrangian to the full covariant form, we find that the functions that gave vanishing a FRW contribution to the Lagrangian were in fact total derivatives in the covariant form, just as in the original fab-four construction \cite{Charmousis:2011bf,charmousis2012self}.  The remaining four non-trivial FLRW contribuitons can be  expressed covariantly as
\\
\begin{equation}\label{v1 lagrangian}
\mathcal{L}_{V_{1}}\big\vert_{FLRW}=-\frac{1}{6}\bar{a}^{3}V_{1}R\big\vert_{FRW},\qquad\qquad\qquad\qquad\quad
\end{equation}
\\
\begin{equation}\label{v3 lagrangian}
\mathcal{L}_{V_{3}}\big\vert_{FLRW}= \frac{1}{16}\bar{a}^{3}V_{3}\hat{G}\big\vert_{FRW},\qquad\qquad\qquad\qquad\quad\;
\end{equation}
\\
\begin{equation}\label{f4 lagrangian}
\mathcal{L}_{f_{4}}\big\vert_{FLRW}=\bar{a}^{3}2f_{4}G^{\mu\nu}\nabla_{\mu}\phi\nabla_{\nu}\phi\big\vert_{FRW},\qquad\qquad\quad\;
\end{equation}
\\
\begin{equation}\label{g5 lagrangian}
\mathcal{L}_{g_{5}}\big\vert_{FLRW}=-2\bar{a}^{3}g_{5}P^{\mu\nu\alpha\beta}\nabla_{\mu}\phi\nabla_{\alpha}\phi\nabla_{\nu}\nabla_{\beta}\phi\big\vert_{FRW},
\end{equation}
\\
where $R$ is the scalar curvature, $\hat{G}=R^{2}-4R^{\mu\nu}R_{\mu\nu}+R^{\mu\nu\rho\lambda}R_{\mu\nu\rho\lambda}$ is the Gauss-Bonnet combination, $G^{\mu\nu}$ is the Einstein tensor and 
\mbox{$P^{\mu\nu\alpha\beta}=-\frac{1}{4}\varepsilon^{\mu\nu\lambda\sigma}R_{\lambda\sigma\gamma\delta}\varepsilon^{\gamma\delta\alpha\beta}$} is the double dual of the Riemann tensor.

Note that the curvature terms contained in these component Lagrangians are evaluated in the Jordan frame, however in this particular case the Jordan and Horndeski frames coincide. Comparing these 
covariant expressions with those found in the Fab-Four \cite{Charmousis:2011bf,charmousis2012self} we see that they are indeed the component Lagrangians that constitute the Fab-Four, as required.  
\\

\subsection{Investigating the conformally coupled case}

Another check of our system of equations  \eqref{X,Y relations 1}, \eqref{X,Y relations 2} is to set $\bar{A}=\bar{A}(\phi)$ and $B=0$, which is equivalent to a conformal transformation. Given that the Horndeski Lagrangian maintains its form under such transformations  then in fact the original calculation of \cite{Charmousis:2011bf,charmousis2012self} actually also includes the case where matter is minimally coupled not to the Horndeski metric, but to a conformally related one, as long as the conformal factor depends on $\phi$, but not $X$. What this means is that if we set $\bar{A}=\bar{A}(\phi)$ and $B=0$ we should recover the Fab-Four, but where the curvature terms in the Lagrangian are expressed using the conformally scaled metric. We find that this is indeed the case.

\subsection{The most general disformal case}

Now that we have confirmed the consistency of our disformal generalisation of the Fab-Four we wish to study the disformal properties of the theory, in other words, we would like to study the effects of
\textit{``switching on''} the disformal part of \eqref{disformal transformation}. Given that a \textit{special} disformal 
transformation \cite{bettoni2014disformal} ($\bar{A}=\bar{A}(\phi)$, $\bar{B}=\bar{B}(\phi)$) does not change the form of the Horndeski action, in order to provide any generalisation beyond Fab-Four theory we need to analyse 
the most general case in which $\bar{A}$ and $\bar{B}$ (and, implicitly, $N$) are in principle dependent on both $\phi$ and $\bar{X}=\bar{X}(\phi,X)$. As we shall see, requiring that the theory adheres to the self-tuning conditions of section \ref{self-tuning constraints} leads to important contraints on the general form of the disformal transformation \eqref{disformal transformation}.

An initial observation is that the functions $X_{i}$ and $Y_{i}$ on the left-hand side (LHS) of the equations given in \eqref{X,Y relations 2} depend on $\phi$ and $\dot{\phi}$, however, the 
right-hand side (RHS) of each of these equations contains terms proportional to $\dot{\bar{A}}(\phi,\bar{X})$, which in this most general case will be dependent on $\phi$, $\dot{\phi}$ \textit{and} $\ddot{\phi}$. 
Therefore in order for $LHS=RHS$ we require that the sum of terms proportional to powers in $\ddot{\phi}$ must vanish in each case. Concentrating on the $X_{0}$ equation (first equation given in 
\eqref{X,Y relations 2}), this means we require that 
\\
\begin{equation}\label{self-tuning barA}
 \alpha(\phi,\dot{\phi})\dot{\bar{A}}(\phi,\bar{X})+\beta(\phi,\dot{\phi})\dot{\bar{A}}^{2}(\phi,\bar{X})+\gamma(\phi,\dot{\phi})\dot{\bar{A}}^{3}(\phi,\bar{X})=0,
\end{equation}
\\
where $\alpha$, $\beta$ and $\gamma$ are the coefficients of $\dot{\bar{A}}$, $\dot{\bar{A}}^{2}$ and $\dot{\bar{A}}^{3}$ in the $X_0$ equation respectively, and are functions of $\phi$ and $\dot{\phi}$.

Expanding $\dot{\bar{A}}(\phi,\bar{X})$ we find that it can be expressed in the form
\\
\begin{equation}\label{barA time derivative}
 \dot{\bar{A}}(\phi,\bar{X})=\dot{\phi}\bar{A}_{,\phi}+\dot{\bar{X}}\bar{A}_{,\bar{X}}=\lambda(\phi,\dot{\phi})+\sigma(\phi,\dot{\phi})\ddot{\phi},
\end{equation}
\\
where we have noted from \eqref{kinetic-term Jordan} that  $\dot{\bar{X}}=\left(\frac{1}{2}\dot{\phi}^{2}\right)^{\small\bullet}=\dot{\phi}\ddot{\phi}$, and also that $\bar{A}_{,\phi}$ and $\bar{A}_{,\bar{X}}$ will be, at most, functions of $\phi$ and $\dot{\phi}$.

Thus, upon inserting \eqref{barA time derivative} into \eqref{self-tuning barA} we arrive at the following equation
\\
\begin{equation}\label{phiddot eq.}
 \left(\alpha\lambda +\beta\lambda^{2}+\gamma\lambda^{3}\right)+\left(\alpha\sigma +2\beta\lambda\sigma +3\gamma\lambda^{2}\sigma\right)\ddot{\phi}
+\left(\beta\sigma^{2}+3\gamma\lambda\sigma^{2}\right)\ddot{\phi}^{2}+\gamma\sigma^{3}\ddot{\phi}^{3}=0,
\end{equation}
\\
Now, as we are assuming that $\bar{A}$ is (in general) non-trivially dependent on both $\phi$ and $\bar{X}$, i.e. $\bar{A}=\bar{A}(\phi,\bar{X})$, it must be that $\bar{A}_{,\phi}\neq 0$ and $\bar{A}_{,\bar{X}}\neq 0$, which from \eqref{barA time derivative}, implies that $\lambda\neq 0$ 
and $\sigma\neq 0$. We require that the coefficient of each power in $\ddot{\phi}$ vanishes, in order for \eqref{phiddot eq.} to hold for all values of $\ddot{\phi}$. Observe the chain of contraints that this 
necessitates  
\\
\begin{align}
 \gamma\sigma^{3}&=0\qquad\Rightarrow\qquad\gamma =0,
 \nonumber\\ \nonumber\\
 \beta\sigma^{2}+3\gamma\lambda\sigma^{2}=\beta\sigma^{2}&=0\qquad\Rightarrow\qquad\beta =0,\nonumber\\ \nonumber\\
 \alpha\sigma +2\beta\lambda\sigma +3\gamma\lambda^{2}\sigma =\alpha\sigma &=0\qquad\Rightarrow\qquad\alpha =0.
\end{align}
\\
and from this we see that the final term in \eqref{phiddot eq.}, $\left(\alpha\lambda +\beta\lambda^{2}+\gamma\lambda^{3}\right)$, is trivially zero. Applying these results to our original expression \eqref{self-tuning barA}, it follows 
that the coefficient for each power in $\dot{\bar{A}}$ must vanish identically. The same argument can then be applied to the remaining equations given in \eqref{X,Y relations 2} to conclude that this result holds for each equation.

The implication of this result is a non-trivial one. Indeed, it is found that by assuming that $\bar{A}$ is a function of both $\phi$ and $X$ and applying the ensuing contraints leads to a contradiction. We therefore conclude from this that in order for the theory to be self-tuning, $\bar{A}$ can be a function of $\phi$ at most, i.e. $\bar{A}=\bar{A}(\phi)$. This is a powerful 
result as the form of the Horndeski Lagrangian does not change under $\phi$-dependent conformal transformations, meaning that if $\hat{g}_{\mu\nu}=A^{2}(\phi)g_{\mu\nu}$, we find
\\
\begin{align}
\mathscr{L}&=\mathscr{L}_{H}(g,\phi,X)+\mathscr{L}_{m}(\bar{g},\psi_{i})\nonumber\\ \nonumber\\ 
&=\tilde{\mathscr{L}}_{H}(\hat{g},\phi,X)+\mathscr{L}_{m}(\hat{g}+\hat{B}^{2}\partial\phi\partial\phi,\psi_{i})\nonumber\\ \nonumber\\
&=\tilde{\mathscr{L}}_{H}(\hat{g},\phi,X)+\mathscr{L}_{m}(\bar{\hat g},\psi_{i}),
\end{align}
\\
where $B$ has been redefined such that $\hat{B}=AB$, and $\hat A=1$.

Thus, given that  $\mathscr{L}_{H}$ maintains its form under conformal transformations, along with the self-tuning requirement, implies that we can effectively set $A=1$ which, due to the relation between the disformal transformation \eqref{disformal transformation} and its inverse \eqref{inverse disformal transformation}, further implies that upon moving to the Jordan frame (as in the previous analysis) we can also effectively set $\bar{A}=1$ 
(\textit{c.f.} \eqref{barA barB constraints}). 

Indeed, given this freedom to set $\bar{A}=1$ it is found that, upon several integration-by-parts, the Horndeski Lagrangian evaluated on an FLRW background $\mathcal{L}_{_{FLRW}}$ can be expressed as 
follows
\\
\begin{equation}\label{general disformal lagrangian}
\begin{split}
 \mathcal{L}_{_{FLRW}}=&\bar{a}^{3}\biggl[N\sqrt{2X}V'_{1}-2V_{1}\frac{s}{\bar{a}}+\left(3\tilde{G}_{5}-3\sqrt{2X}G_{5}-V_{2}\right)\left(\frac{s}{\bar{a}}\right)^{2}\biggr]
\biggl[\bar{H}-\frac{s}{\bar{a}}\biggr]\\ \\
&+\bar{a}^{3}\biggl[N\sqrt{2X}V'_{2}+6NG_{4}-3N\sqrt{2X}\tilde{G}_{5,\phi}+2V_{1}\biggr]\biggl[\bar{H}^{2}-\left(\frac{s}{\bar{a}}\right)^{2}\biggr]\\ \\
&+2\bar{a}^{3}\frac{X\sqrt{2X}}{N^{2}}G_{5,X}\biggl[\bar{H}^{3}-\left(\frac{s}{\bar{a}}\right)^{3}\biggr],
\end{split}
\end{equation}
\\
\\
where $\dot{\phi}=N\sqrt{2X}$, $N(t)=\frac{1}{\sqrt{1-2B^2X}}$ (using (\ref{barA barB constraints}), (\ref{kinetic-term Einstein}) and (\ref{transformed factors})), $V'_{i}\equiv\frac{dV_{i}}{d\phi}$, $(\;\;)_{,\phi}\equiv\frac{\partial}{\partial \phi}(\;\;)$, and 
$(\;\;)_{,X}\equiv\frac{\partial}{\partial X}(\;\;)$.
\newpage
Accordingly, if we can solve the corresponding set of equations  \eqref{X,Y relations 1} and \eqref{X,Y relations 2} with $\bar A=1$
\begin{align}
\label{X,Y general disformal}
&K-\sqrt{2X}\tilde{G}_{3,\phi}=\sqrt{2X}V'_{0}(\phi),\nonumber\\ \nonumber\\
&3\sqrt{2X}G_{3}-3\tilde{G}_{3}-6\sqrt{2X}G_{4,\phi}=N\sqrt{2X}V'_{1}(\phi)+3V_{0}(\phi),\nonumber\\ \\
&\frac{12}{N}XG_{4,X}-\frac{6}{N}G_{4}-\frac{6}{N}XG_{5,\phi}-6G_{4}+3N\sqrt{2X}\tilde{G}_{5,\phi}=N\sqrt{2X}V'_{2}(\phi)+2V_{1}(\phi),\nonumber\\ \nonumber\\
&\frac{2}{N^{2}}X\sqrt{2X}G_{5,X}-3\sqrt{2X}G_{5}+3\tilde{G}_{5}=N\sqrt{2X}V'_{3}(\phi)+V_{2}(\phi).\nonumber
\end{align} 
then the theory is guaranteed to be self-tuning\footnote{Note that solving this system with $N=1$ gives the Fab-Four case.}. Clearly, as these equations involve an unknown function of $X$, ${B}(\phi,X)$, expressed in the above set as an unkown function $N$, we cannot integrate these in general, but rather must do so by first specifying the arbitrary function. Another approach would be to specify the various $G_i$, $K$, and solve for what $N$ has to be. Even having done so, one would then like to be able to write down the covariant form of the Lagrangian, which may not be an easy task.

\section{A particular solution for the most general disformal case}

In this section we shall present a simple solution to the set of differential equations \eqref{X,Y general disformal}, and in doing so explicitly show that the corresponding Lagrangian can not be put into Fab-Four form. Before proceeding we would like to remind the reader of a few of the equations that have been used earlier in the paper as they shall be employed heavily in this analysis. The first we draw attention to is the set of differential equations \eqref{X,Y general disformal}, derived in the previous subsection, from which one can in principle determine the forms of the functions $K,G_{3},G_{4}$ and $G_{5}$ once the disformal coupling has been given in the form of $B(\phi,X)$, or equivalently \footnote{From \eqref{barA barB constraints}, \eqref{kinetic-term Einstein} and \eqref{transformed factors} we see that \mbox{$N=\left[ 1-2B^2X\right]^{-1/2}$}} $N(t)$. Futhermore, we shall be using the kinetic term \eqref{kinetic-term Einstein} for the scalar field $\phi$, along with the auxilliary functions, $\tilde{G}_{3}$ and $\tilde{G}_{5}$, both of which are defined in \eqref{aux functions} in terms of $G_{3}$ and $G_{5}$.
We first note from \eqref{X,Y general disformal} that we have a system of four differential equations with five unknown functions, $N,K,G_{3},G_{4}$ and $G_{5}$. The first equation in \eqref{X,Y general disformal} is, of course, trivial to solve (for $K$) leaving us with three remaining equations and four unknown functions. Thus, whichever way we look at it, our system is \emph{under-determined}, which is what allows us to choose how matter is to be disformally coupled. 
It is important to note that one cannot choose $N$ to be of the form $N=f(\phi)\left(2X\right)^{-1/2}$, the reason being that we require from \eqref{kinetic-term Einstein} that $N\sqrt{2X}=\dot{\phi}$, hence if $N$ were of this form one would arrive at inconsistent solutions since $\phi$ and $\dot{\phi}$ are independent variables. 

Having set-up our preliminary framework we now proceed to solve the differential equation for $G_{5}$ (fourth and final equation in \eqref{X,Y general disformal}). To keep matters simple we shall make the following choice for $N$
\begin{equation}\label{G5 def}
 \frac{2}{N^{2}}X\sqrt{2X}G_{5,X}\equiv N\sqrt{2X}V'_{3}.
\end{equation}
Given this, and using our relation between $G_{5}$ and $\tilde{G}_{5}$ \eqref{aux functions} we are left with the following differential equation  
\begin{equation}
 2X\tilde{G}_{5,X}-\tilde{G}_{5}=-\frac{V_{2}}{3}
\end{equation}
whose solution is given by $\tilde{G}_{5}=f(\phi)\sqrt{2X}+\frac{1}{3}V_{2}$ and thus implying that 
\begin{equation}\label{G5 disformal}
 G_{5}(\phi,X)=f(\phi).
\end{equation}
Using \eqref{G5 def}, this leads to $N\sqrt{2X}V'_{3}=0$, which we solve with
\begin{equation}
V_{3}=const,
\end{equation}
as we take $N$ to be non-vanishing. We therefore see that in fact \eqref{G5 def} gives no constraint on the form of $N$.

We now turn our attention to $G_{4}$ which can be determined from the third equation in \eqref{X,Y general disformal}. Again, in the interest of obtaining an analytic solution, we note that as the integration function $f(\phi)$ is arbitrary we restrict to the case $f=const$, and in doing so we find that 
\begin{equation} 
 12XG_{4,X}-6\left(1+N\right)G_{4}=2NV_{1},
\end{equation}
at which point we make our choice of $N$ to be
\begin{equation}
N(t)=g(\phi)X-1,
\end{equation}
yielding
\begin{equation}\label{G4 diff eq disformal}
12XG_{4,X}-6g(\phi)XG_{4}=2NV_{1}.
\end{equation}
Using that the function $V_{1}(\phi)$ is still free we further simplify restrict to $V_{1}=0$, giving the following solution to \eqref{G4 diff eq disformal}
\begin{equation} \label{G4 disformal}
 G_{4}(\phi,X)=h(\phi)\exp{\left(\frac{g(\phi)X}{2}\right)}.
\end{equation}
This leads us onto to the penultimate differential equation to be solved, the second equation of \eqref{X,Y general disformal}, which we simplify by taking $g(\phi)=g=const.$ and $h(\phi)=h=const.$ We now have that $G_{4,\phi}=0$ and so one finds
\begin{equation}
 6X\tilde{G}_{3,X}-3\tilde{G}_{3}=3V_{0}\qquad\Rightarrow\qquad\tilde{G}_{3}(\phi,X)=I(\phi)\sqrt{2X}-V_{0}(\phi)
\end{equation}
where $I(\phi)$ is an arbitrary integration function. (note that we have made use of the relation between $G_{3}$ and $\tilde{G}_{3}$ (\emph{c.f.} \eqref{aux functions}), and taken into account from earlier that we set $V_{1}=0$). Hence,
\begin{equation}\label{G3 disformal}
 G_{3}(\phi,X)=I(\phi).
\end{equation}
Finally, we can trivially determine the functional form of $K$ from the remaining equation in \eqref{X,Y general disformal},
\begin{equation}\label{K disformal}
 K(\phi,X)=2I(\phi)X.
\end{equation}
In summary then, we find the following solution for the set of functions
\begin{align}
 G_{5}(\phi,X)&=f,\nonumber \\  
 G_{4}(\phi,X)&=h\exp{\left(\frac{gX}{2}\right)},\\ \nonumber
 G_{3}(\phi,X)&=I(\phi),\\ \nonumber\\ \nonumber
 K(\phi,X)&=2I(\phi)X.
\end{align}
Upon inserting these into our disformally self-tuning Lagrangian \eqref{general disformal lagrangian}, we find that for this particular scenario it has the following form
\\
\begin{equation}\label{disformal l analysis}
 \mathcal{L}_{_{FRW}}=6\bar{a}^{3}Nhe^{\left(gX/2\right)}\biggl[\bar{H}^{2}-\left(\frac{s}{\bar{a}}\right)^{2}\biggr],\quad N=gX-1.
\end{equation}
\\
This is an important result as it shows explicitly that the system of coupled differential equations \eqref{X,Y general disformal} is solvable and leads to non-trivial (consistent) results, even for a simplified case as was analysed here. The analysis of this particular case has also identified which choices of $N$ are not permissible if one is to obtain a consistent solution set. Furthermore, whilst the Lagrangian \eqref{disformal l analysis} is clearly of self-tuning form \eqref{self-tuning Lagrangian}, it cannot be put into Fab-Four form \cite{charmousis2012self}, highlighting the fact that in the most general disformal case our theory extends beyond Fab-Four.

\section{Summary \& outlook}

We have been able to show that it is indeed possible to generalise the Fab-Four theory and obtain a self-tuning theory of gravity in which the self-tuning scalar field $\phi$ is disformally coupled to matter. It has been shown that this generalisation 
is consistent with known results, reproducing the Fab-Four theory for both minimal coupling to the Horndeski metric and a minimal coupling to a Weyl-rescaled Horndeski metric - as long as the scaling function depends on $\phi$ but not $X$. Furthermore, we have found that the requirement that the scalar field $\phi$ is able to self-tune, and thus screen the cosmological constant, places strong constraints on any form of disformal coupling to matter in the theory. Indeed, it was found that, in general, the conformal part of any disformal coupling to matter necessarily must be a function of $\phi$ alone (as opposed to being a function of both $\phi$ and its canonical kinetic term $X$). Given this result, the general disformal case can be simplified by effectively setting the conformal function $A(\phi)$ to unity (due to the Horndeski Lagrangian maintaining its form under Weyl rescaling by a function of $\phi$), and it is subsequently found that the theory can automatically be expressed in a self-tuning form in general. Thus if one can determine the Horndeski functions $K(\phi,X)$, $G_{i}(\phi,X)$ ($i=3,4,5$), then the theory is guaranteed to be 
self-tuning. The caveat of this result is that the differential equations that must be solved in order to determine the Horndeski functions can not be solved in general, only on a case-by-case basis, owing to the presence of an arbitrary function. We have, however, been able to solve the system for a particular case with non-trivial disformal coupling. This simple analysis also provided information on inadmissible choices of the lapse function $N(t)$, and served to highlight that the resulting Lagrangian cannot be expressed in Fab-Four form.

Throughout this paper all analysis was carried out by evaluating the theory on an FLRW background, as such the theory, in its current form, is not covariant. Ideally, the aim would be to find a covariant form of the theory however at present there does not appear to be an ``obvious'' approach to take in 
accomplishing this task (we are unable to utilise the same procedure as in the Fab-Four case due to the additional terms introduced by a disformal coupling). It is possible that future research into this area may uncover an analogue approach to that taken in deriving a 
covariant form of the Fab-Four. Indeed, further analysis of the results presented in this paper may enable one to formulate a covariant expression using the particular case of the form of the Lagrangian on an FLRW background as a starting point.

Finally, it is worth noting that there has been a recent gain in interest into the possiblity of extending beyond Horndeski theory (see, for example, \cite{Gleyzes:2014dya,lin2014hamiltonian,gleyzes2015exploring,kobayashi2015breaking,gao2014unifying,gao2014hamiltonian}), and given that the results obtained for the general disformal case cannot be expressed in Fab-Four form these may well prove useful as a starting point for such an extension.

\acknowledgments

WE and PM are funded by STFC. In particular, WE and PM would like to thank Antonio Padilla for his helpful discussions on the intricacies of the cosmological constant problem and issues encountered whilst constructing such a generalisation of the Fab-Four.

\appendix
\section{Proof that $\mathscr{H}= \tilde{\mathscr{H}}$ and $\varepsilon^{\phi}= \tilde{\varepsilon^{\phi}}$} 

Here we shall explicitly prove that relations \eqref{general self-tuning} are correct, consequently implying that in order for the self-tuning conditions in section \ref{self-tuning constraints} to 
be satisfied, our putative Lagrangian \eqref{self-tuning Lagrangian} is indeed necessary and thus provides a general description of the self-tuning theory.

We start by considering two Horndeski theories defined by \eqref{full l z form} and \eqref{self-tuning Lagrangian}, satisfying the condition \eqref{generic-on-shell}. 
Focusing on the Hamiltonian 
constraints first considering the Hamiltonian, we note that in principle $\mathscr{H}$ and $\tilde{\mathscr{H}}$ differ by a function 
$\Delta\mathscr{H}=\Delta\mathscr{H}(\bar{a},\dot{\bar{a}},\phi,\dot{\phi},\ddot{\phi})$, i.e.
\begin{equation}
\mathscr{H}= \tilde{\mathscr{H}}+\Delta\mathscr{H}
\end{equation}
[The functional dependence of $\Delta\mathscr{H}$ arises from the fact that matter couples in the same way in both Horndeski-like theories (by assumption)].

If we now go (generically) on-shell, then
\begin{align}
\mathscr{H}+\rho_{m}&= \left(\tilde{\mathscr{H}}+\Delta\mathscr{H}\right) +\rho_{m}\nonumber\\ \nonumber\\
&=\left(\tilde{\mathscr{H}}+\rho_{m}\right)+\Delta\mathscr{H} \nonumber\\ \nonumber\\
&=\Delta\mathscr{H}=0
\end{align}
where we have noted that the expression in the brackets on the second line vanishes by virtue of the on-shell requirements \eqref{generic-on-shell}. Hence we 
find that $\Delta\mathscr{H}$ should vanish when the on-shell conditions are satisfied, however, we are yet to determine whether $\Delta\mathscr{H}$ vanishes
algebraically or identically.

Now, by assumption, $\Delta\mathscr{H}$ cannot depend on $\rho_{m}$, as if it did then this would imply that $\mathscr{H}$ also depends on $\rho_{m}$, however, 
$\mathscr{H}$ generates the time evolution of the scalar field, $\phi$, which (again by assumption) does not \emph{directly} couple to the matter sector, and hence $\mathscr{H}$
should be independent of $\rho_{m}$ (as otherwise this would imply that $\phi$ interacts directly with matter). Consequently, $\Delta\mathscr{H}$ cannot vanish by virtue
of the equation $\tilde{\mathscr{H}}=-\rho_{m}$. In cases $I)$ and $II)$ ($Z_{i}$ non-linear in $\ddot{\phi}$ and linear in $\ddot{\phi}$, respectively, \emph{c.f.} end of section \ref{Zi cases 1}), we see 
that $\tilde{\varepsilon}^{\phi}$ (\emph{c.f.} \eqref{scalar eom}) contains $\dddot{\bar{a}}$, however $\tilde{\mathscr{H}}$ (\emph{c.f.} \eqref{Jordan frame Hamiltonian}) 
does not and so we cannot use $\tilde{\varepsilon}^{\phi}=0$ to enforce $\Delta\mathscr{H}=0$ (as in both cases there would be non-trivial terms remaining with no corresponding 
term to cancel with). In case $III)$ ($Z_{i}$ independent of $\ddot{\phi}$, \emph{c.f.} end of section \ref{Zi cases 1}), we see that $\tilde{\varepsilon}^{\phi}$ contains $\ddot{\bar{a}}$, however $\tilde{\mathscr{H}}$ 
does not and so we cannot use $\tilde{\varepsilon}^{\phi}=0$ to enforce $\Delta\mathscr{H}=0$ in this case either. Hence, as we cannot use the dynamical equations 
$\tilde{\mathscr{H}}=-\rho_{m}$ and $\tilde{\varepsilon}^{\phi}=0$ to enforce $\Delta\mathscr{H}=0$ we are forced to conclude that $\Delta\mathscr{H}$ is 
identically zero. In other words,
\begin{equation}
\mathscr{H}=\tilde{\mathscr{H}} 
\end{equation}
\\ 
We now turn our attention to the scalar equation of motion, $\varepsilon^{\phi}$ \eqref{scalar eom}. The analysis in this case is a little more involved than for the Hamiltonian case,
and to aid ourselves we first consider the following. 

As was the case with the Hamiltonian, in principle, the Lagrangians of our putative self-tuning theory and a general self-tuning theory will differ by a function
$\Delta Z_{i}=\Delta Z_{i}(\bar{a},\phi,\dot{\phi},\ddot{\phi})= Z_{i}-\tilde{Z}_{i}$ $\bigl[$as $\Delta\mathcal{L}=\mathcal{L}-\tilde{\mathcal{L}}=
\bar{a}^{3}\sum_{i=0}^{3}\left(Z_{i}-\tilde{Z}_{i}\right)\bar{H}^{i}=\bar{a}^{3}\sum_{i=0}^{3}\Delta Z_{i}\bar{H}^{i}\bigr]$. Given this, we claim that 
\begin{equation} \label{delta Z}
\Delta Z_{i}=\sigma_{i}(\bar{a},\phi)\dot{\phi}^{1-i} 
\end{equation}
To prove this we refer to our earlier derivation of $\mathscr{H}$, \eqref{Jordan frame Hamiltonian}, from which we can infer that 
\begin{align} \label{GH}
&\Delta\mathscr{H}= \bar{a}^{3}\sum_{i=0}^{3}\Biggl[\left(i-1\right)\Delta Z_{i}\bar{H}^{i}+\dot{\phi}\Delta Z_{i,\dot{\phi}}\bar{H}^{i}
+\left(i-3\right)\dot{\phi}\Delta Z_{i,\ddot{\phi}}\bar{H}^{i+1}-i\dot{\phi}\Delta Z_{i,\ddot{\phi}}\frac{\ddot{\bar{a}}}{\bar{a}}\bar{H}^{i-1} 
\nonumber\\ 
&\qquad\qquad\qquad\qquad-\dot{\phi}^{2}\Delta Z_{i, \phi , \ddot{\phi}}\bar{H}^{i}-\dot{\phi}\ddot{\phi}\Delta Z_{i, \dot{\phi} , \ddot{\phi}}\bar{H}^{i}
-\bar{a}\dot{\phi}\Delta Z_{i, \bar{a} , \ddot{\phi}}\bar{H}^{i+1}+\ddot{\phi}\Delta Z_{i,\ddot{\phi}}\bar{H}^{i}
-\dot{\phi}\dddot{\phi}\Delta Z_{i, \ddot{\phi} , \ddot{\phi}}\bar{H}^{i}\Biggr]\nonumber\\ \nonumber\\ &\qquad=0
\end{align}
We now know that when we are (generically) on-shell $\Delta\mathscr{H}$ vanishes identically and this implies that each of the terms in the above equation vanish 
individually. By equating powers in $\bar{H}$ we immediately see that 
\begin{equation}
i\dot{\phi}\Delta Z_{i,\ddot{\phi}}\frac{\ddot{\bar{a}}}{\bar{a}}=0\;\;\Rightarrow\;\; \Delta Z_{i,\ddot{\phi}}=0
\end{equation}
and this holds whatever dependence $Z_{i}$ has on $\ddot{\phi}$. Hence, we find that 
\begin{equation}\label{differential Z}
\Delta\mathscr{H}= \bar{a}^{3}\sum_{i=0}^{3}\Biggl[\left(i-1\right)\Delta Z_{i}+\dot{\phi}\Delta Z_{i,\dot{\phi}}\Biggr]\bar{H}^{i} =0\;\;\Rightarrow\;\;
\left(i-1\right)\Delta Z_{i}+\dot{\phi}\Delta Z_{i,\dot{\phi}}=0
\end{equation}
which leaves us with a first-order differential equation for $\Delta Z_{i}$, and as $\Delta Z_{i,\ddot{\phi}}=0$ we can infer that it is a function of 
$\bar{a},\;\phi$ and $\dot{\phi}$, at most, i.e. $\Delta Z_{i}=\Delta Z_{i}(\bar{a},\phi,\dot{\phi})$. Upon integrating \eqref{differential Z} with respect to $\dot{\phi}$ we find
\begin{align}
\label{z difference}
&\ln (\Delta Z_{i})= \left(1-i\right)\ln (\dot{\phi})+ f_{i}(\bar{a},\phi) \nonumber\\ \nonumber\\
&\Rightarrow\;\;\Delta Z_{i}=\sigma_{i}(\bar{a},\phi)\dot{\phi}^{1-i}
\end{align}
where $\sigma_{i}(\bar{a},\phi)$ is an arbitrary function of $\bar{a}$ and $\phi$.

Now that we are equipped with this additional information we shall proceed with our analysis of the scalar equation of motion. As was the case for the Hamiltonian, 
in principle, the scalar equation of motion, $\tilde{\varepsilon^{\phi}}$, for our putative self-tuning theory will differ from that of a general self-tuning 
theory, $\varepsilon^{\phi}$, by a function $\Delta\varepsilon^{\phi}=\Delta\varepsilon^{\phi}(\bar{a},\dot{\bar{a}},\ddot{\bar{a}},\phi,\dot{\phi},\ddot{\phi})$ 
as follows
\begin{equation}
\varepsilon^{\phi}= \tilde{\varepsilon^{\phi}}+\Delta\varepsilon^{\phi}
\end{equation}
Now, as $\varepsilon^{\phi}$ describes the motion of the scalar field, $\phi$, which by assumption does not directly couple to the matter sector, it therefore cannot depend 
on $\rho_{m}$ (for the same reasons as discussed in the Hamiltonian case). Consequently, this implies that $\Delta\varepsilon^{\phi}$ should be independent of 
$\rho_{m}$ also. Thus, when on-shell, $\Delta\varepsilon^{\phi}$ cannot vanish by virtue of the equation $\tilde{\mathscr{H}}=-\rho_{m}$. Noting from the set-up to 
this analysis, that $\Delta\mathcal{L}=\bar{a}^{3}\sum_{i=0}^{3}\Delta Z_{i}\bar{H}^{i}$, we have
\begin{align}
\label{eom difference}
\Delta\varepsilon^{\phi}&=\frac{\partial(\Delta\mathcal{L})}{\partial\phi}-\frac{d}{dt}\left[\frac{\partial(\Delta\mathcal{L})}{\partial\dot{\phi}}\right]
+\frac{d^{2}}{dt^{2}}\left[\frac{\partial(\Delta\mathcal{L})}{\partial\ddot{\phi}}\right] \nonumber\\ \nonumber\\
&=\sum_{i=0}^{3}\biggl[\bar{a}^{3}\Delta Z_{i,\phi}\bar{H}^{i}-\frac{d}{dt}\left(\bar{a}^{3}\Delta Z_{i,\dot{\phi}}\bar{H}^{i}\right)
+\frac{d^{2}}{dt^{2}}\left(\bar{a}^{3}\Delta Z_{i,\ddot{\phi}}\bar{H}^{i}\right)\biggr]
\end{align}
However, we know from previously, that regardless of the dependence of $Z_{i}$ on $\ddot{\phi}$, $\Delta Z_{i,\ddot{\phi}}=0$ and so    
\begin{align}
\Delta\varepsilon^{\phi}&=\sum_{i=0}^{3}\biggl[\bar{a}^{3}\Delta Z_{i,\phi}\bar{H}^{i}
-\frac{d}{dt}\left(\bar{a}^{3}\Delta Z_{i,\dot{\phi}}\bar{H}^{i}\right)\biggr]
 \nonumber\\ \nonumber\\
&=\bar{a}^{3}\sum_{i=0}^{3}\Biggl[\dot{\phi}\Delta Z_{i,\phi, \dot{\phi}}\bar{H}^{i}-\ddot{\phi}\Delta Z_{i,\dot{\phi}, \dot{\phi}}\bar{H}^{i}
-\left(3-i\right)\Delta Z_{i,\dot{\phi}}\bar{H}^{i+1}-\bar{a}\Delta Z_{i,\bar{a},\dot{\phi}}\bar{H}^{i+1} \nonumber\\
&\qquad\qquad\quad-i\Delta Z_{i,\dot{\phi}}\frac{\ddot{\bar{a}}}{\bar{a}}\bar{H}^{i-1}+\Delta Z_{i,\phi}\bar{H}^{i}\Biggr]
\end{align}
Now, when (generically) on-shell we require that the conditions \eqref{generic-on-shell} are satisfied, and this implies that, on-shell, 
$\Delta\varepsilon^{\phi}=0$. We know that this condition can not be satisfied by virtue of the equation $\tilde{\mathscr{H}}=-\rho_{m}$, and so, at best, it 
vanishes by virtue of the equation $\tilde{\varepsilon}^{\phi}=0$.

For cases $I)$ and $II)$ ($Z_{i}$ non-linear in $\ddot{\phi}$ and linear in $\ddot{\phi}$, respectively, \emph{c.f.} end of section \ref{Zi cases 1}), $\tilde{\varepsilon}^{\phi}$ contains $\dddot{\bar{a}}$ 
and so we cannot use it to substitute in for $\ddot{\bar{a}}$ in $\Delta\varepsilon^{\phi}$ (as there will be no corresponding term in $\Delta\varepsilon^{\phi}$ 
to cancel out the $\dddot{\bar{a}}$ term introduced in such a substitution). As such, in these cases, $\Delta\varepsilon^{\phi}$ must vanish identically (i.e. 
each term in \eqref{eom difference} must vanish individually). Accordingly, through equating powers in $\bar{H}$ this implies that
\begin{equation}
i\Delta Z_{i,\dot{\phi}}\frac{\ddot{\bar{a}}}{\bar{a}}=0\;\;\Rightarrow\;\; i\Delta Z_{i,\dot{\phi}}=0
\end{equation}
and using \eqref{z difference},
\begin{equation}
i\Delta Z_{i,\dot{\phi}}=i\left(1-i\right)\sigma_{i}\dot{\phi}^{-i}=0 
\end{equation}
For $i=0,1$ we see that the left-hand side vanishes due to the term $i\left(1-i\right)$ and so permits a non-trivial form for $\sigma_{i}$. However, for $i=2,3$, 
we see that $i\left(1-i\right)\neq 0$ and so we are forced to conclude that 
\begin{align}
\sigma_{2}&=0 \\ \nonumber\\ 
\sigma_{3}&=0 
\end{align}
For the case in which $Z_{i}$ is independent of $\ddot{\phi}$ (case $III$, \emph{c.f.} end of section \ref{Zi cases 1}), $\tilde{\varepsilon}^{\phi}$ does not contain $\dddot{\bar{a}}$, but it does still 
contain $\ddot{\bar{a}}$ and so we must be more careful in our analysis (as, in principle, this could be substituted in to $\Delta\varepsilon^{\phi}$ such that 
the terms cancel algebraically such that $\Delta\varepsilon^{\phi}$ is not identically zero, but $\Delta\varepsilon^{\phi}=0$ is satisfied). To proceed, we note 
that for $Z_{i,\ddot{\phi}}=0$, the scalar equation of motion has the form
\begin{equation}
\varepsilon^{\phi}=A(\bar{a},\dot{\bar{a}},\phi,\dot{\phi})\ddot{\bar{a}} +B(\bar{a},\dot{\bar{a}},\phi,\dot{\phi})\ddot{\phi}
+C(\bar{a},\dot{\bar{a}},\phi,\dot{\phi})
\end{equation}
and similarly for $\tilde{\varepsilon}^{\phi}$ (with ``tilded'' functions $\tilde{A}$, $\tilde{B}$, $\tilde{C}$ replacing the functions $A$, $B$ and $C$), as in 
both cases the coupling to matter is the same. It follows then, that 
\begin{equation}
\ddot{\bar{a}}=\frac{1}{\tilde{A}}\left[\tilde{\varepsilon}^{\phi}-\tilde{B}\ddot{\phi}-\tilde{C}\right] 
\end{equation}
which leads us to the expression
\begin{align}
\label{delta eom}
\Delta\varepsilon^{\phi}&= \varepsilon^{\phi}-\tilde{\varepsilon}^{\phi}=\left[A-\tilde{A}\right]\ddot{\bar{a}}+\left[B-\tilde{B}\right]\ddot{\phi}
+\left[C-\tilde{C}\right] \nonumber\\ \nonumber\\
&=\frac{\Delta A}{\tilde{A}}\left[\tilde{\varepsilon}^{\phi}-\tilde{B}\ddot{\phi}-\tilde{C}\right]+\Delta B\ddot{\phi} +\Delta C \nonumber\\ \nonumber\\
&=\frac{\Delta A}{\tilde{A}}\tilde{\varepsilon}^{\phi}+\frac{\tilde{A}\Delta B-\tilde{B}\Delta A}{\tilde{A}}\ddot{\phi}+\frac{\tilde{A}\Delta C
-\tilde{C}\Delta A}{\tilde{A}}
\end{align}
where $\Delta A= A-\tilde{A}$ and similarly for $\Delta B$ and $\Delta C$.

Now, $\Delta\varepsilon^{\phi}$ ought to vanish by virtue of the equation $\tilde{\varepsilon}^{\phi}=0$ when on-shell, and so we can immediately infer that
\begin{equation}
\Delta\varepsilon^{\phi}= \frac{\Delta A}{\tilde{A}}\tilde{\varepsilon}^{\phi}
\end{equation}
This is because the second term (in the third line) on the right-hand side of \eqref{delta eom} contains $\ddot{\phi}$ whereas the third term does not, and hence
they cannot cancel one another out (even in principle). Therefore it must also be the case that 
\begin{equation}
\tilde{A}\Delta B=\tilde{B}\Delta A\; , \qquad \tilde{A}\Delta C=\tilde{C}\Delta A
\end{equation}
Furthermore, upon comparison of our expressions for $\Delta\varepsilon^{\phi}$ we can infer that 
\begin{align}
&\Delta A=-i\Delta Z_{i,\dot{\phi}}\frac{\bar{H}^{i-1}}{\bar{a}}\bar{a}^{3}\;\;\Rightarrow\;\;
\tilde{A}=-i\tilde{Z}_{i,\dot{\phi}}\frac{\bar{H}^{i-1}}{\bar{a}}\bar{a}^{3}\\ \nonumber\\
&\Delta B=-\Delta Z_{i,\dot{\phi},\dot{\phi}}\bar{H}^{i}\bar{a}^{3}\;\;\Rightarrow\;\;\tilde{B}=-\tilde{Z}_{i,\dot{\phi},\dot{\phi}}\bar{H}^{i}\bar{a}^{3} 
\end{align}
and from \eqref{z difference} 
\begin{align}
\Delta\varepsilon^{\phi}&=\sum_{i=0}^{3}\left[\bar{a}^{3}\Delta Z_{i,\phi}\bar{H}^{i}-\frac{d}{dt}\left(\bar{a}^{3}\Delta Z_{i,\dot{\phi}}\bar{H}^{i}\right)\right]
\nonumber\\ \nonumber\\
&= \bar{a}^{3}\sum_{i=0}^{3}\biggl[\sigma_{i,\phi}\dot{\phi}^{1-i}\bar{H}^{i}-3\left(1-i\right)\sigma_{i}\dot{\phi}^{-i}\bar{H}^{i+1}
-i\left(1-i\right)\sigma_{i}\dot{\phi}^{-i}\frac{\ddot{\bar{a}}}{\bar{a}}\bar{H}^{i-1}+i\left(1-i\right)\sigma_{i}\dot{\phi}^{-i}\bar{H}^{i+1}\nonumber\\
&\qquad\qquad\qquad-\left(1-i\right)\sigma_{i,\phi}\dot{\phi}^{1-i}\bar{H}^{i}+i\left(1-i\right)\sigma_{i}\dot{\phi}^{-(1+i)}\ddot{\phi}\bar{H}^{i}
-\left(1-i\right)\bar{a}\sigma_{i,\bar{a}}\dot{\phi}^{-i}\bar{H}^{i+1}\biggr]
\end{align}
Hence,
\begin{equation}
\Delta A=-i\left(1-i\right)\sigma_{i}\frac{\bar{H}^{i-1}}{\dot{\phi}^{i}}\bar{a}^{2}\;,\qquad\Delta 
B=-\frac{\bar{a}\bar{H}}{\dot{\phi}}\left[-i\left(1-i\right)\sigma_{i}\frac{\bar{H}^{i-1}}{\dot{\phi}^{i}}\bar{a}^{2}\right]= 
-\frac{\bar{a}\bar{H}}{\dot{\phi}}\Delta A
\end{equation}
Now, we require that $\tilde{A}\Delta B=\tilde{B}\Delta A$ in order for $\Delta\varepsilon^{\phi}$ to vanish by virtue of $\tilde{\varepsilon}^{\phi}=0$, and so 
utilising the above relations.
\begin{equation}
\tilde{A}\Delta B= -\tilde{A}\frac{\bar{a}\bar{H}}{\dot{\phi}}\Delta A=\tilde{B}\Delta A
\end{equation}
Assuming that $\Delta A\neq 0$ (i.e. $\Delta\varepsilon^{\phi}$ vanishes by virtue of $\tilde{\varepsilon}^{\phi}=0$ and not identically) this gives
\begin{align}
-\bar{a}\bar{H}\tilde{A}=\dot{\phi}\tilde{B}\;\;&\Rightarrow\;\; -\bar{a}\bar{H}\left[-i\tilde{Z}_{i,\dot{\phi}}\frac{\bar{H}^{i-1}}{\bar{a}}\bar{a}^{3}\right]
=\dot{\phi}\left[-\tilde{Z}_{i,\dot{\phi},\dot{\phi}}\bar{H}^{i}\bar{a}^{3}\right] \nonumber\\ \nonumber\\
&\Rightarrow i\tilde{Z}_{i,\dot{\phi}}\bar{H}^{i}=-\dot{\phi}\tilde{Z}_{i,\dot{\phi},\dot{\phi}}\bar{H}^{i}\;\;\Rightarrow\;\;
\frac{\tilde{Z}_{i,\dot{\phi},\dot{\phi}}}{\tilde{Z}_{i,\dot{\phi}}}=-i\frac{1}{\dot{\phi}}\nonumber\\ \nonumber\\ 
&\Rightarrow \ln(\tilde{Z}_{i,\dot{\phi}})= -i\ln(\dot{\phi})+f_{i}\left(\bar{a},\phi\right)\nonumber\\ \nonumber\\ 
&\Rightarrow\tilde{Z}_{i,\dot{\phi}}=\alpha_{i}\left(\bar{a},\phi\right)\dot{\phi}^{-i}
\end{align}
where $\alpha_{i}\left(\bar{a},\phi\right)$ is an arbitrary `constant' of integration (with respect to $\dot{\phi}$).

It is evident from this expression that 
\begin{align}
&\text{For $i\neq 1$ :}\quad\tilde{Z}_{i,\dot{\phi}}=\alpha_{i}\left(\bar{a},\phi\right)\dot{\phi}^{-i}\;\;\Rightarrow\;\; 
\tilde{Z}_{i}\left(\bar{a},\phi,\dot{\phi}\right)=u_{i}\left(\bar{a},\phi\right)\dot{\phi}^{1-i}+v_{i}\left(\bar{a},\phi\right)\nonumber\\ \nonumber\\ 
&\text{For $i= 1$ :}\quad\tilde{Z}_{1,\dot{\phi}}=\alpha_{1}\left(\bar{a},\phi\right)\dot{\phi}^{-1}\;\;\Rightarrow\;\; 
\tilde{Z}_{1}\left(\bar{a},\phi,\dot{\phi}\right)=u_{1}\left(\bar{a},\phi\right)\ln(\dot{\phi})+v_{1}\left(\bar{a},\phi\right)
\end{align}
(where, as $u_{i}$ and $v_{i}$ are arbitrary functions we have absorbed any additional terms, introduced through integrating, into them).

Accordingly, $\tilde{Z}_{i}$ has the following form(s) for each value of $i=0,1,2,3$:
\begin{equation}
\tilde{Z}_{i}\left(\bar{a},\phi,\dot{\phi}\right) =\Biggr\lbrace\begin{matrix} u_{1}\left(\bar{a},\phi\right)\ln(\dot{\phi})+v_{1}\left(\bar{a},\phi\right) &\qquad
\text{for $i=1$}\\u_{i}\left(\bar{a},\phi\right)\dot{\phi}^{1-i}+v_{i}\left(\bar{a},\phi\right)&\qquad\text{for $i\neq 1$}\end{matrix}
\end{equation}
Notice, however, from \eqref{self-tuning Lagrangian} that $\tilde{\mathcal{L}}$ vanishes when \emph{on-shell-in-$\bar{a}$} which leads to the following
\begin{align}
\tilde{\mathcal{L}}_{k}=0\;\;&\Rightarrow\;\; 0=\tilde{Z}_{i}\left(\frac{s}{\bar{a}}\right)^{i}\nonumber\\ \nonumber\\ 
&\Rightarrow\tilde{Z}_{i}=0
\end{align}
as, in general, $\frac{s}{\bar{a}}\neq 0$. In patricular, this implies that $u_{i}=0 \Rightarrow \tilde{Z}_{i,\dot{\phi}}=0$. 

Now, this scenario is highly undesirable as it leads to a highly constrained trivial theory in which the only solution permitted is a Minkowski spacetime (in direct violation of the self-tuning filter \emph{c.f.} section \ref{self-tuning constraints}. We are 
therefore forced to conclude that for a non-trivial theory, in actual fact, $\Delta A=0$ and hence $\Delta\varepsilon^{\phi}$ vanishes identically (as in cases 
$I$ and $II$). This result implies that 
\begin{equation}
-i\left(1-i\right)\sigma_{i}\frac{\bar{H}^{i-1}}{\dot{\phi}^{i}}\bar{a}^{2}=0 
\end{equation}
and we see that for $i=0,1$ this condition is satisfied by $i\left(1-i\right)$, however, for $i=2,3$ we see that $i\left(1-i\right)\neq 0$ and, as such, we 
conclude that 
\begin{equation}
\sigma_{2}=0\;,\qquad\sigma_{3}=0 
\end{equation}
Therefore, in all three cases, $\sigma_{2}=0=\sigma_{3}$ and $\Delta\varepsilon^{\phi}$ vanishes identically, i.e. $\varepsilon^{\phi}= \tilde{\varepsilon^{\phi}}$.

\section{Derivation of system of differential equations for $K(\phi,X)$, $G_{i}(\phi,X)$ ($i=3,4,5$)}

Given the expression found for $\Delta Z_{i}$ \eqref{delta Z} in \emph{appendix A} and the requirement that $\Delta\varepsilon^{\phi}=0$ identically we now know that $\sigma_{2}=0=\sigma_{3}$ 
in all three cases (\emph{c.f.} end of section \ref{Zi cases 1}) and so we aim to determine a more explicit form for the remaining non-trivial functions $\sigma_{0}$ and $\sigma_{1}$. To this end, note that 
$\Delta Z_{i,\ddot{\phi}}=0$ in all three cases, and further that $\Delta Z_{i}$ takes the form given in \eqref{z difference}. As such, 
\begin{align}
\Delta\varepsilon^{\phi}&=\sum_{i=0}^{3}\biggl[\bar{a}^{3}\Delta Z_{i,\phi}\bar{H}^{i}-\frac{d}{dt}\left(\bar{a}^{3}\Delta Z_{i,\dot{\phi}}\bar{H}^{i}\right)
\biggr] \nonumber\\ \nonumber\\ 
&=\bar{a}^{3}\sum_{i=0}^{3}\Biggl[\dot{\phi}\Delta Z_{i,\phi, \dot{\phi}}\bar{H}^{i}-\ddot{\phi}\Delta Z_{i,\dot{\phi}, \dot{\phi}}\bar{H}^{i}
-\left(3-i\right)\Delta Z_{i,\dot{\phi}}\bar{H}^{i+1}-\bar{a}\Delta Z_{i,\bar{a},\dot{\phi}}\bar{H}^{i+1} \nonumber\\
&\qquad\qquad\quad-i\Delta Z_{i,\dot{\phi}}\frac{\ddot{\bar{a}}}{\bar{a}}\bar{H}^{i-1}+\Delta Z_{i,\phi}\bar{H}^{i}\Biggr]
\end{align}
We now know that $\Delta\varepsilon^{\phi}$ must vanish on-shell and this immediately implies that
\begin{align}
&i\Delta Z_{i,\dot{\phi}}\frac{\ddot{\bar{a}}}{\bar{a}}=0 \\ \nonumber\\ 
& \ddot{\phi}\Delta Z_{i,\dot{\phi}, \dot{\phi}}=0
\end{align}
Furthermore, we now know that $\Delta\varepsilon^{\phi}$ must vanish identically and so the remaining terms must also vanish
\begin{align}
\dot{\phi}\Delta Z_{i,\phi, \dot{\phi}}\bar{H}^{i}-\left(3-i\right)\Delta Z_{i,\dot{\phi}}\bar{H}^{i+1}-\bar{a}\Delta Z_{i,\bar{a},\dot{\phi}}\bar{H}^{i+1} 
+\Delta Z_{i,\phi}\bar{H}^{i}&=0 \nonumber\\ \nonumber\\ 
\Rightarrow\;\;\left[i\dot{\phi}\sigma_{i,\phi}-\left(1-i\right)\left(\left(3-i\right)\sigma_{i}+\bar{a}\sigma_{i,\bar{a}}\right)\bar{H}\right]
\frac{\bar{H}^{i}}{\dot{\phi}^{i}}&=0
\end{align}
Thus, through equating powers in $\bar{H}$, we can infer from this that
\begin{equation}
\sigma_{1,\phi}=3\sigma_{0}+\bar{a}\sigma_{0,\bar{a}}=\frac{1}{\bar{a}^{2}}\left[3\bar{a}^{2}\sigma_{0}+\bar{a}^{3}\sigma_{0,\bar{a}}\right]
= \frac{1}{\bar{a}^{2}}\left(\bar{a}^{3}\sigma_{0}\right)_{,\bar{a}}\quad\Rightarrow\quad\bar{a}^{2}\sigma_{1,\phi}=\left(\bar{a}^{3}\sigma_{0}\right)_{,\bar{a}}
\end{equation}
and hence, by defining a function $\mu = \mu (\bar{a},\phi)$, we can unambiguously express $\sigma_{0}$ and $\sigma_{1}$ in the following forms 
\begin{equation}\label{sigmas explicit}
\bar{a}^{2}\sigma_{1}(\bar{a},\phi)= \mu_{,\bar{a}}\;,\qquad \bar{a}^{3}\sigma_{0}(\bar{a},\phi)= \mu_{,\phi}
\end{equation}
Given the analysis thus far we now claim that our self-tuning ansatz, $\tilde{\mathcal{L}}$ and the general self-tuning Lagrangian, 
$\mathcal{L}$, differ by a total derivative, i.e.
\begin{equation}
\Delta\mathcal{L}=\mathcal{L}-\tilde{\mathcal{L}}=\frac{d}{dt}\bigg(\mu (\bar{a},\phi)\bigg) 
\end{equation}
To prove this claim we note that $\Delta Z_{i}$ has the form \eqref{z difference} and that $\sigma_{2}=0=\sigma_{3}$, and as such
\begin{equation}
\Delta Z_{0}=\sigma_{0}\dot{\phi}\;,\qquad \Delta Z_{1}=\sigma_{1}\;,\qquad \Delta Z_{2}=0=\Delta Z_{3}
\end{equation}
From this we can deduce that 
\begin{equation}
\Delta\mathcal{L}=\bar{a}^{3}\sum_{i=0}^{3}\Delta Z_{i}\bar{H}^{i}= \bar{a}^{3}\left[\Delta Z_{0}+\Delta Z_{1}\bar{H}\right]
=\bar{a}^{3}\left[\sigma_{0}\dot{\phi}+\sigma_{1}\bar{H}^{1}\right]
\end{equation}
and upon noting the forms of $\sigma_{0}$ and $\sigma_{1}$, \eqref{sigmas explicit}, we can re-express this as
\begin{equation}
\Delta\mathcal{L}=\bar{a}^{3}\sigma_{0}\dot{\phi}+\bar{a}^{3}\sigma_{1}\bar{H}= \dot{\phi}\mu_{,\phi}+\dot{\bar{a}}\mu_{,\bar{a}}
=\frac{d}{dt}\bigg(\mu (\bar{a},\phi)\bigg) 
\end{equation}
as required.

Given the functional expression for the Lagrangian, \eqref{full l}, we observe that the functions $Z_{i}$ can be expressed in the following form
\begin{equation}
\label{appendix X and Y}
Z_{i}=X_{i}-\frac{k}{\bar{a}^{2}}Y_{i}=X_{i}+\frac{s^{2}}{\bar{a}^{2}}Y_{i} 
\end{equation}
where $s \equiv\sqrt{-k}$ ($=\dot{\bar{a}}$ when \emph{on-shell-in-$\bar{a}$}), and in particular, we note that $Y_{2}=0=Y_{3}$.

Our aim now is to determine how $X_{i}$ and $Y_{i}$ are related and their functional forms. We start from our knowledge that the Lagrangian of the general self-
tuning theory and our ansatz differ by a total derivative, i.e. $\mathcal{L}=\tilde{\mathcal{L}}+\frac{d}{dt}\bigg(\mu (\bar{a},\phi)\bigg)$, and hence, on the 
right-hand side we have that 
\begin{align}
\tilde{\mathcal{L}}+\frac{d}{dt}\bigg(\mu (\bar{a},\phi)\bigg)&= \bar{a}^{3}\sum_{i=1}^{3}\tilde{Z}_{i}\biggl[\bar{H}^{i}-\left(\frac{s}{\bar{a}}\right)^{i}\biggr]
+\dot{\phi}\mu_{,\phi}+\dot{\bar{a}}\mu_{,\bar{a}}\nonumber\\ \nonumber\\
&=\bar{a}^{3}\Biggl[\tilde{Z}_{1}\left[\bar{H}-\left(\frac{s}{\bar{a}}\right)\right]+\tilde{Z}_{2}\left[\bar{H}^{2}-\left(\frac{s}{\bar{a}}\right)^{2}\right]+
\tilde{Z}_{3}\left[\bar{H}^{3}-\left(\frac{s}{\bar{a}}\right)^{3}\right]+\dot{\phi}\mu_{,\phi}+\dot{\bar{a}}\mu_{,\bar{a}}\Biggr]
\end{align}
whilst on the left-hand side we have that
\begin{align}
\mathcal{L} &=\bar{a}^{3}\sum_{i=1}^{3}Z_{i}\bar{H}^{i}=\bar{a}^{3}\sum_{i=1}^{3}\left[X_{i}+\frac{s^{2}}{\bar{a}^{2}}Y_{i} \right]\bar{H}^{i}
\nonumber\\ \nonumber\\
&=\bar{a}^{3}\Biggl[X_{0}+\frac{s^{2}}{\bar{a}^{2}}Y_{0}+\left[X_{1}+\frac{s^{2}}{\bar{a}^{2}}Y_{1}\right]\bar{H}+X_{2}\bar{H}^{2}+X_{3}\bar{H}^{3}\Biggr]
\end{align}
If we now equate powers in $\bar{H}$ we obtain the following set of equations
\begin{align}
-\frac{s}{\bar{a}}\tilde{Z}_{1}-\left(\frac{s}{\bar{a}}\right)^{2}\tilde{Z}_{2}-\left(\frac{s}{\bar{a}}\right)^{3}\tilde{Z}_{3}+\bar{a}^{-3}\dot{\phi}\mu_{,\phi}
&=X_{0}+\frac{s^{2}}{\bar{a}^{2}}Y_{0}\qquad\qquad\qquad\qquad\qquad\qquad\nonumber\\ \nonumber\\
\tilde{Z}_{1}+\bar{a}^{-2}\mu_{,\bar{a}}&=X_{1}+\frac{s^{2}}{\bar{a}^{2}}Y_{1}\nonumber\\ \nonumber\\
\tilde{Z}_{2}&=X_{2}\nonumber\\ \nonumber\\
\tilde{Z}_{3}&=X_{3}
\end{align}
Upon substituting the coefficients of $\bar{H}$, $\bar{H}^{2}$ and $\bar{H}^{3}$ into the coefficient of $\bar{H}^{0}$ we find that 
\begin{align}
-\frac{s}{\bar{a}}\left[-\bar{a}^{-2}\mu_{,\bar{a}} + X_{1}+\frac{s^{2}}{\bar{a}^{2}}Y_{1}\right]-\left(\frac{s}{\bar{a}}\right)^{2}X_{2}
-\left(\frac{s}{\bar{a}}\right)^{3}X_{3}+\bar{a}^{-3}\dot{\phi}\mu_{,\phi}&=X_{0}+\frac{s^{2}}{\bar{a}^{2}}Y_{0} \nonumber\\ \nonumber\\
\Rightarrow\;\;\bar{a}^{-3}\dot{\phi}\mu_{,\phi}+s\bar{a}^{-3}\mu_{,\bar{a}}-X_{0}-\frac{s}{\bar{a}}X_{1}
-\left(\frac{s}{\bar{a}}\right)^{2}\left[X_{2}+Y_{0}\right]
-\left(\frac{s}{\bar{a}}\right)^{3}\left[X_{3}+Y_{1}\right]&=0
\end{align}
To analyse this further we shall first expand $\bar{a}^{-3}\mu (\bar{a},\phi)$ as a power series, in terms if $\frac{s}{\bar{a}}$, around $s=0$ in the following 
manner 
\begin{equation}
\bar{a}^{-3}\mu (\bar{a},\phi)=\sum_{i}V_{i}\left(\phi\right)\left(\frac{s}{\bar{a}}\right)^{i} 
\end{equation}
where $V_{i}\left(\phi\right)$ are (as of yet) arbitrary functions of the scalar field. We therefore have that
\begin{equation}
\dot{\phi}\sum_{i}V'_{i}\left(\phi\right)\left(\frac{s}{\bar{a}}\right)^{i}+\sum_{i}\left(3-i\right)V_{i}\left(\phi\right)\left(\frac{s}{\bar{a}}\right)^{i+1}
-X_{0}-\frac{s}{\bar{a}}X_{1}-\left(\frac{s}{\bar{a}}\right)^{2}\left[X_{2}+Y_{0}\right]-\left(\frac{s}{\bar{a}}\right)^{3}\left[X_{3}+Y_{1}\right]=0 
\end{equation}
(where $V'_{i}(\phi)\equiv\frac{dV_{i}}{d\phi}$ and similarly for higher order derivatives).

Hence, equating powers in $\frac{s}{\bar{a}}$ leads to the following relations
\begin{align}
\label{appendix X,Y relations 1}
X_{0}&= \text{const.} + \dot{\phi}V'_{0}\left(\phi\right)\\ \nonumber\\
X_{1}&=\dot{\phi}V'_{1}\left(\phi\right)+3V_{0}\left(\phi\right)\\ \nonumber\\
X_{2}+Y_{0}&=\dot{\phi}V'_{2}\left(\phi\right)+2V_{1}\left(\phi\right)\\ \nonumber\\
X_{3}+Y_{1}&=\dot{\phi}V'_{3}\left(\phi\right)+V_{2}\left(\phi\right)
\end{align}
Using \eqref{appendix X and Y} we can now compare with the general Horndeski Lagrangian (whose form we calculated earlier), \eqref{full l}, to determine the forms of 
$X_{i}$ and $Y_{i}$ and/or their relations to one another.


\end{document}